\documentclass[prb,reprint,showpacs,groupedaddress]{revtex4-1}

\usepackage{graphicx}
\usepackage[colorlinks, citecolor=blue]{hyperref}
\usepackage{amsmath}
\usepackage{amssymb}
\begin{document}

\title{Enhancing superconducting critical current by randomness}

\author{Y. L. Wang$^{1,2}$}\email{ylwang@anl.gov or ywang35@nd.edu}
\author{L. R. Thoutam$^{1,3}$}
\author{Z. L. Xiao$^{1,3}$}\email{xiao@anl.gov or zxiao@niu.edu}
\author{B. Shen$^{1}$}
\author{J. E. Pearson$^{1}$}
\author{R. Divan$^{4}$}
\author{L. E. Ocola$^{4}$}
\author{G. W. Crabtree$^{1,5}$}
\author{W. K. Kwok$^{1}$}

\affiliation{$^{1}$Materials Science Division, Argonne National Laboratory, Argonne, Illinois 60439, USA}

\affiliation{$^{2}$Department of Physics, University of Notre Dame, Notre Dame, Indiana 46556, USA}

\affiliation{$^{3}$Department of Physics, Northern Illinois University, DeKalb, Illinois 60115, USA}

\affiliation{$^{4}$Center for Nanoscale Materials, Argonne National Laboratory, Argonne, Illinois 60439, USA}

\affiliation{$^{5}$Departments of Physics, Electrical and Mechanical Engineering, University of Illinois at Chicago, Illinois 60607, USA}

\date{\today}

\begin{abstract}
The key ingredient of high critical currents in a type-II superconductor is defect sites that ‘pin’ vortices.  Contrary to earlier understanding on nano-patterned artificial pinning, here we show unequivocally the advantages of a random pinscape over an ordered array in a wide magnetic field range. We reveal that the better performance of a random pinscape is due to the variation of its \textit{local density of pinning sites} (LDOPS), which mitigates the motion of vortices. This is confirmed by achieving even higher enhancement of the critical current through a conformally mapped random pinscape, where the distribution of the LDOPS is further enlarged. The demonstrated key role of LDOPS in enhancing superconducting critical currents gets at the heart of random versus commensurate pinning. Our findings highlight the importance of random pinscapes in enhancing the superconducting critical currents of applied superconductors.
\end{abstract}


\maketitle
\section{Introduction}
In a type-II superconductor which constitutes most applied superconductors, above a certain lower critical field, the magnetic flux will penetrate into the superconductor and form quantized vortices, each consisting of exactly one quantum of flux surrounded by circulating supercurrents. The motion of these vortices driven by the Lorentz force induced by an applied current dissipates energy and limits the potential applications of superconductors. In order to overcome this limitation, intensive efforts have been made to immobilize or 'pin' the vortices.\cite{Ref1,Ref2,Ref3,Ref4,Dam1999Ref5,Bugoslavsky2001Ref6,Ref5,Ref6,Ref10,Ref14,Ref9,Ref11,CompositeAntidots,Ref17,Ref18,Ref25,Ref26,PRBBitterDecoration,VortexIcePRLXiao,ConformalWang,ConformalAPL,VortexIceNatureNanoTech,PRB2012MaskIrradiationPeroidic,Fang2013Fef16,Miura2013Ref17}
Among them, artificial pinning sites introduced into superconducting thin films through nano-patterning have been widely investigated. A periodic pinning array,\cite{PRL1995SquareMagnetization,Ref9,Ref11,CompositeAntidots,Ref17,Ref18,New21,PRB1998PeroidReichhardt,PRB2001PeroidReichhardt,PRB2001PeroidicReichhardt2} can significantly enhance the critical current when the vortex lattice is commensurate with the array. However, the enhancement effect is greatly reduced at magnetic fields away from the commensurate matching fields. In order to overcome this shortcoming, many complex arrangements, such as Penrose lattice arrays\cite{Ref20,Ref21,Ref22,Ref23,Ref24,Ref27}, honeycomb arrays\cite{HoneycombJAP2005,Honeycomb2007Reichhardt,Honeycomb2008Reichhardt,Honeycomb2012Xiao}, diluted periodic arrays \cite{DisorderInOrder2009Kemmler,ConformalAPL} and pinscapes with a density gradient \cite{Ref25,Ref26,Ref27,ConformalRayPRL, ConformalWang, ConformalAPL, ConformalRayPRB}, were proposed and investigated for enhancing critical current in magnetic fields.

The above pinscapes, however, are technically challenging to create in large samples, limiting their practical applicability. One system that has been largely overlooked in nano-patterning investigations is the random pinscape, which constitutes an intrinsic disordered system and can be formed during materials synthesis\cite{Ref1,Ref4,Dam1999Ref5,MacManus-Driscoll2004,Miura2013Ref17} or through ex-situ artificial defect formation.\cite{Bugoslavsky2001Ref6,Fang2013Fef16,Krusin-Elbaum2000} Although random pinscapes have been proven to be able to enhance critical current in superconductors,\cite{Bugoslavsky2001Ref6,Fang2013Fef16,MacManus-Driscoll2004,Krusin-Elbaum2000,PhysRevB.82.014509} however, previous investigations have reported undesirable outcomes for random pinscapes as compared to periodic pinscapes: Bitter decoration indicates that a random pinscape is less effective in trapping vortices;\cite{PRBBitterDecoration} transport experiments on Nb films with random pinning holes reveal lower critical currents than even a non-patterned reference film.\cite{Ref14} Computer simulations show that in a random pinscape, the larger interaction energy of the closely spaced vortices pinned in the high density areas of the pinscape reduces the pinning effectiveness and critical current relative to a periodic pinning array.\cite{Random1996Reichhardt,ConformalRayPRL,ConformalRayPRB} 

However, the reported results may be fraught with uncertainties. For example, film degradation from nano-patterning could result in low quality films, rendering the results ambiguous (see section \ref{Considerations} for detailed discussion on this issue). Bitter decoration only probes static vortices. Assumptions made in computer simulations, such as the pinning and vortex interaction energy ratio, may be valid only in a specific range of the experimental parameters.

Here we demonstrate through direct comparison, that pinning effects of randomly distributed nanoscale holes patterned into superconducting films can enhance the superconducting critical currents over a wider magnetic field range than periodic pinning holes. 
Utilizing the Voronoi diagram to quantify the \textit{local density of pinning sites} (LDOPS), we demonstrate that a large spatial variation in LDOPS is the key to enhance the critical current at high magnetic fields. Our finding is further confirmed by achieving even higher critical currents using a pinscape conformally transformed from a random hole-array, which has an additional global hole-density gradient with wider spatial distribution of LDOPS. Although the conformally mapped random pattern has no local ordering compared to a conformal array of a hexagonal hole-lattice which can be commensurate with the vortex lattice, the former yields nearly identical pinning effects as the latter. Since both random and hexagonal conformal pinscapes have similar LDOPS distributions, our results highlight the critical role played by the LDOPS distribution in the enhancement of the critical current at high magnetic fields. 
Our results should stimulate more investigations to reveal the potential capabilities and advantages of tailored random pinscapes in enhancing the superconducting critical current, contributing to the development of superconductors for high current applications.

\begin{figure}[tbh]
 \begin{center}
 \centerline{\includegraphics[width=0.48\textwidth]{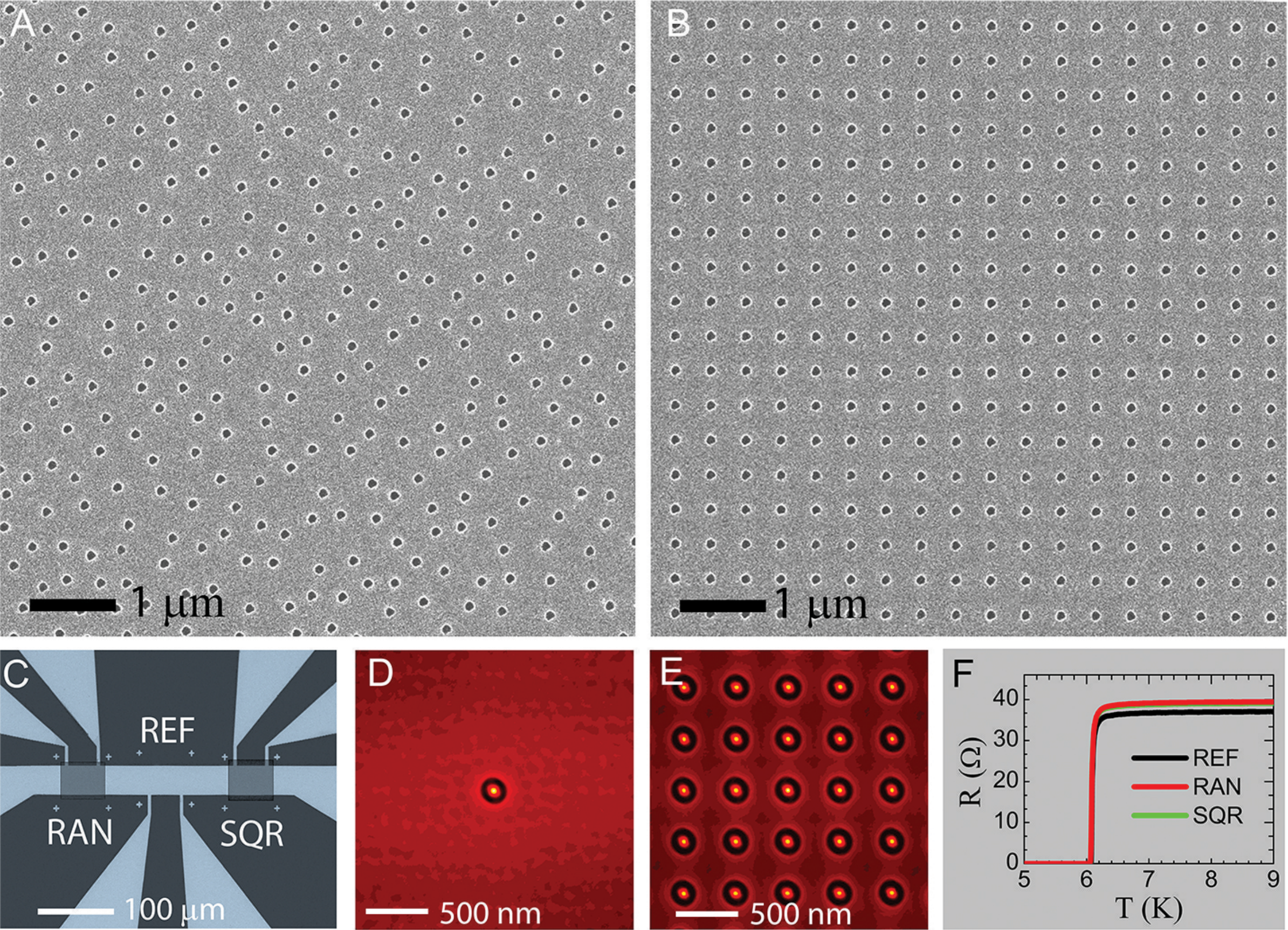}}
 \caption{\label{fig:fig1} (\textbf{A}) and (\textbf{B}) SEM micrographs of two patterned Sections in Sample I. (\textbf{C}) Image of Sample I which contains a unpatterned reference section (Section REF), a section with randomly distributed holes (Section RAN) and a section with a square array of holes (Section SQR). The bright region is MoGe film, black region is Si/SiO$2$ substrate and the two shadowed regions are patterned region on MoGe film. The hole-hole spacing in (B) is $420$ nm. The average densities of the holes in (A) and (B) are identical. (\textbf{D}) and (\textbf{E}) Auto-correlations for SEM micrographs of Sections RAN and SQR, respectively. (\textbf{F}) Superconducting transition curves of the three sections in Sample I.}
\end{center}
 \end{figure}

\section{Sample fabrication and characterization}

The experiments were carried out on Mo$_{0.79}$Ge$_{0.21}$ (MoGe) superconducting thin films with weak intrinsic pinning,\cite{Ref35} enabling transport measurements at temperatures far below the zero-field critical temperature $T_{c0}$ to avoid complications from phenomena originating from the Little-Parks effect.\cite{Ref36,Sochnikov2010,New40} 
Films of $50$ nm thick were sputtered from a MoGe alloy target onto silicon substrates with oxide layer. The samples were first patterned into $50$ $\mu$m wide microbridges containing three sections using photolithography. Different arrays of nanoscale holes with diameter of $100$ nm were introduced through electron-beam lithography followed by reactive ion etching. First, a ZEP resist layer with thickness about $450$ nm was spin-coated on the MoGe film (bridge) deposited on a Si substrate. After baking at $150$ $^o$C for $3$ minutes, it was exposed to electron-beam with a dose of $75$ $\mu$C/cm$^2$ and energy of $30$ keV. Then it was developed in Xylene for $40$ seconds and in IPA for $30$ seconds sequentially. Plasma of sulfur hexafluoride (SF$_6$) at a pressure of $20$ mTorr, a flow rate of $20$ sccm and a power of $50$ W was adopted for etching MoGe. The etching times were $2$ minutes $10$ seconds which produces through holes on $50$ nm thick MoGe film.

Figure \ref{fig:fig1}C shows a picture of our typical sample, which consists of three sections of the same dimensions ($50 \mu$m $\times$ $50 \mu$m) on a $50$ nm thick Mo$_{0.79}$Ge$_{0.21}$ film. The un-patterned Section REF serves as the reference to Sections RAN and SQR, which contain a random array and a square lattice of nanoscale holes with diameter of $100$ nm, respectively, as shown by the scanning electron microscopy (SEM) images in Fig. \ref{fig:fig1}A and B for Sample I. The period of the square lattice of Fig.~\ref{fig:fig1}B is $420$ nm. Since this work focuses on the effect of distribution of the holes on vortex pinning, the average density of holes in the random and square arrays were kept the same. As shown in Fig. \ref{fig:fig1}F for Sample I, the zero magnetic field critical temperatures for all three sections are identical at $6.07$ K, indicating no degradation of the sample quality in the patterning process, which is crucial for a reliable study of pinscapes containing non-uniform hole distributions (see section \ref{Considerations} on reliable comparisons). The slightly larger normal state resistance of the patterned sections, as compared to that of the reference one, is due to the reduction of conducting Mo$_{0.79}$Ge$_{0.21}$ in the hole area. In the random array we set a lower limit of $300$ nm for the center to center distance between holes to avoid any overlap, ensuring that all holes are discrete and identical. Such a treatment does not affect the 'randomness' of the hole distribution, as indicated by the single bright spot in the center of the autocorrelation image shown in Fig.\ref{fig:fig1}D for the holes in Section RAN. For comparison, the autocorrelation image for Section SQR, which has a perfect four-fold symmetry is shown in Fig.\ref{fig:fig1}E.

Transport measurements were carried out using a standard dc four-probe method. The current flows horizontally along the long length of the microbridge. The voltages of the three bridges in each sample were measured at the same time. The applied magnetic field is always perpendicular to the film plane. The critical current was defined with a voltage criterion of $2$ $\mu$V.
\section{Results and discussion}

\subsection{Periodic vs random} 
\begin{figure}[tbh]
   \begin{center}
  \centerline{\includegraphics[width=0.48\textwidth]{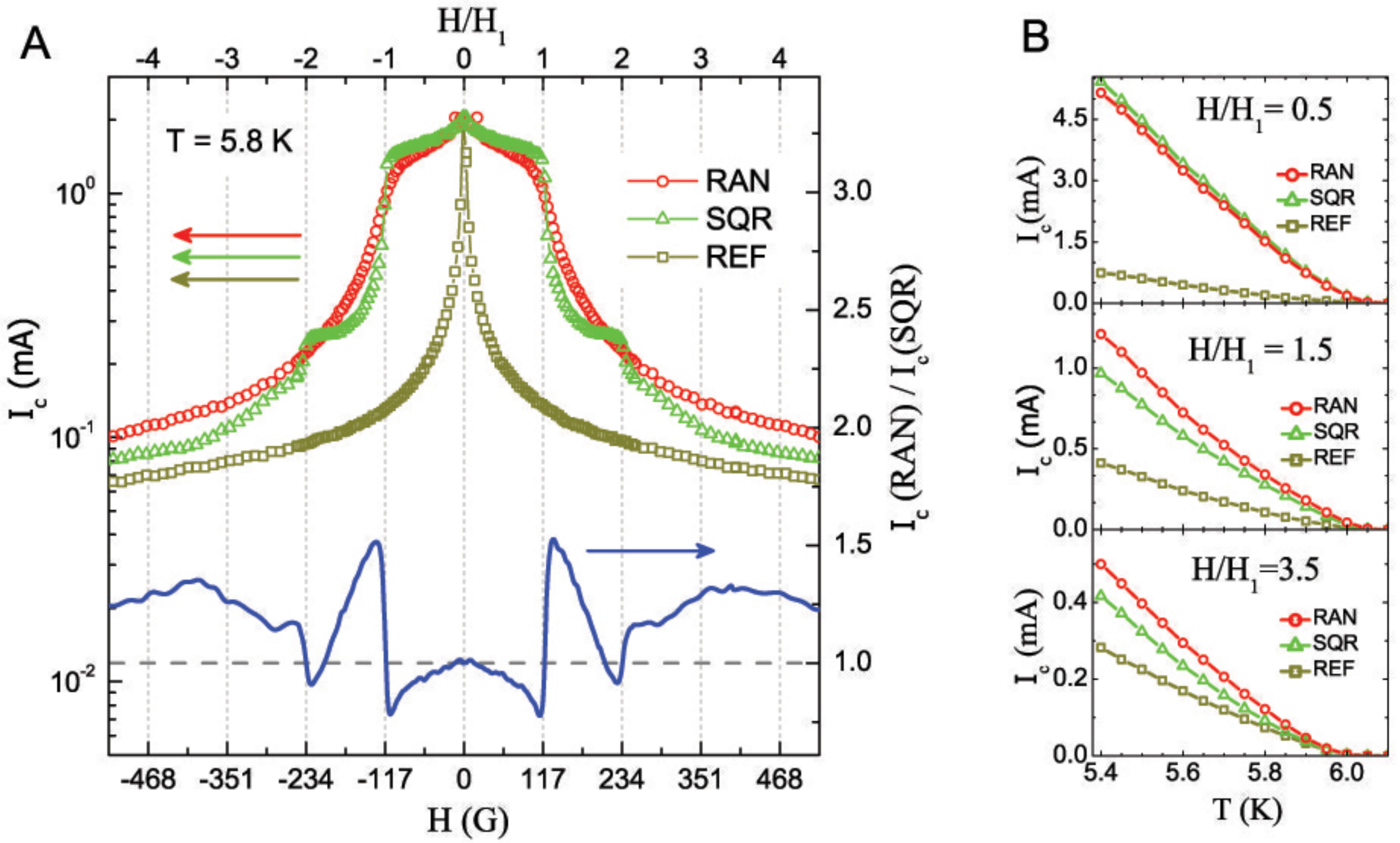}}
  \caption{\label{fig:fig2} (\textbf{A}) Magnetic field dependence of the critical currents. The green line shows the critical current ratio for Section RAN and Section SQR, which indicates Section RAN has higher critical currents over a wider range of high magnetic fields. (\textbf{B}), Temperature dependence of critical currents at magnetic fields of $0.5H_1$,$1.5H_1$ and $3.5H_1$.}
 \end{center}
  \end{figure}

In Fig.\ref{fig:fig2}A we present the magnetic field dependence of the critical current $I_c(H)$ for all three sections of Sample I at temperature of $5.8$ K. It is evident that the introduction of nanoscale holes in Sections RAN and SQR enhances the critical current at all applied magnetic fields. At the calculated first matching field ($H = H_1$) for Section SQR we observed a step in the $I_c(H)$ curve (green triangles). Such a sharp decrease of the critical current at magnetic fields beyond $H_1$, is similar to that observed in perforated Nb films at low temperatures,\cite{Ref11} and is due to the appearance of interstitial vortices that are not directly pinned by the holes.\cite{ConformalRayPRL,Ref37} The steep step of the $I_c(H)$ curve at first matching field indicates each hole with a diameter of $d \approx 100$ nm only allows one vortex to reside in it, which is consistent with the theoretical saturation number $n_{si}=d/4\xi(T)\approx 1$, where $\xi(5.8 K)=24$ nm.\cite{ConformalWang} The smaller step that appears at the second matching field should be due to the caging effect whereby an interstitial vortex is localized by the pinned vortices in the surrounding holes.\cite{ConformalWang,Honeycomb2012Xiao,New44} 
The $I_c(H)$ curve (red circles) for Section RAN shows a smoother decay.

Section RAN has lower $I_c$ than Section SQR at $H<H_1$. This is not only because the randomly distributed pinned vortices have higher interaction energy, but also because a random pinscape traps fewer vortices as compared to its periodic counterpart:\cite{PRBBitterDecoration} in Section SQR all the vortices can be trapped in the holes when $H\leq H_1$. In Section RAN, however, the distribution of LDOPS is non-uniform. In regions with high LDOPS some holes will not be occupied by vortices, while in the region with low LDOPS interstitial vortices exist between the holes even if the global average density of vortices is smaller than that of the holes. Both the higher vortex interaction energy in high LDOPS areas and the existence of interstitial vortices in low LDOPS areas make a random hole-array less effective in vortex pinning, resulting in lower critical currents.

Once the magnetic field exceeds $H_1$, the critical current for Section SQR is greatly suppressed due to the appearance of interstitial vortices that can move in the straight easy-flow channels between the rows of holes. In contrast, in Section RAN the empty holes in the high LDOPS regions will be gradually filled with new vortices and the interstitial vortices will only proliferate in regions with low LDOPS, resulting in a smoother decay of the critical current. Since the global average hole and vortex densities are the same in the two patterned sections, there will be more interstitial vortices in Section RAN than that in Section SQR even at $H>H_1$, a condition that will persist until all holes are filled with vortices, which occurs at high enough field to overcome the high interaction energy of vortices pinned in the high LDOPS regions.
Thus, the observed higher critical currents in Section RAN clearly indicate that a random pinscape is more effective in preventing the motion of interstitial vortices. In contrast, the interstitial vortices are effectively pinned by the square hole-arrays at magnetic fields near the second matching field through caging effects,\cite{New44} resulting in higher critical currents over a very narrow field range about the matching fields. Overall, as indicated by the green line representing the ratio of critical current for Section RAN and SQR in Fig. \ref{fig:fig2}A, a random pinning array can perform better in enhancing critical currents over a wider range of magnetic fields as compared to a square lattice of pinning sites by mitigating the motion of interstitial vortices. The temperature dependence of the critical currents shown in Fig. \ref{fig:fig2}B indicates that the above conclusions are valid at all experimentally accessible temperatures.

The behavior of a random pinscape shown in Fig. \ref{fig:fig2}A is very similar to that of a conformally mapped pinscape from a hexagonal hole-array, which also outperforms a periodic pinscape beyond the first matching field.\cite{ConformalWang} The better performance of the conformal array is attributed to both the global density gradient of pinning sites and the local hexagonal ordering that is commensurate with the vortex lattice.\cite{ConformalRayPRL} However, neither of these advantages exists in our random pinscape that has a uniform global hole density. The common feature in the random and conformal pinscapes is the variation in the LDOPS. As elaborating below, we will show that a wide distribution of LDOPS is the key for achieving high critical currents over a wide range of magnetic fields.

 \begin{figure}[tbh]
  \begin{center}
  \centerline{\includegraphics[width=0.48\textwidth]{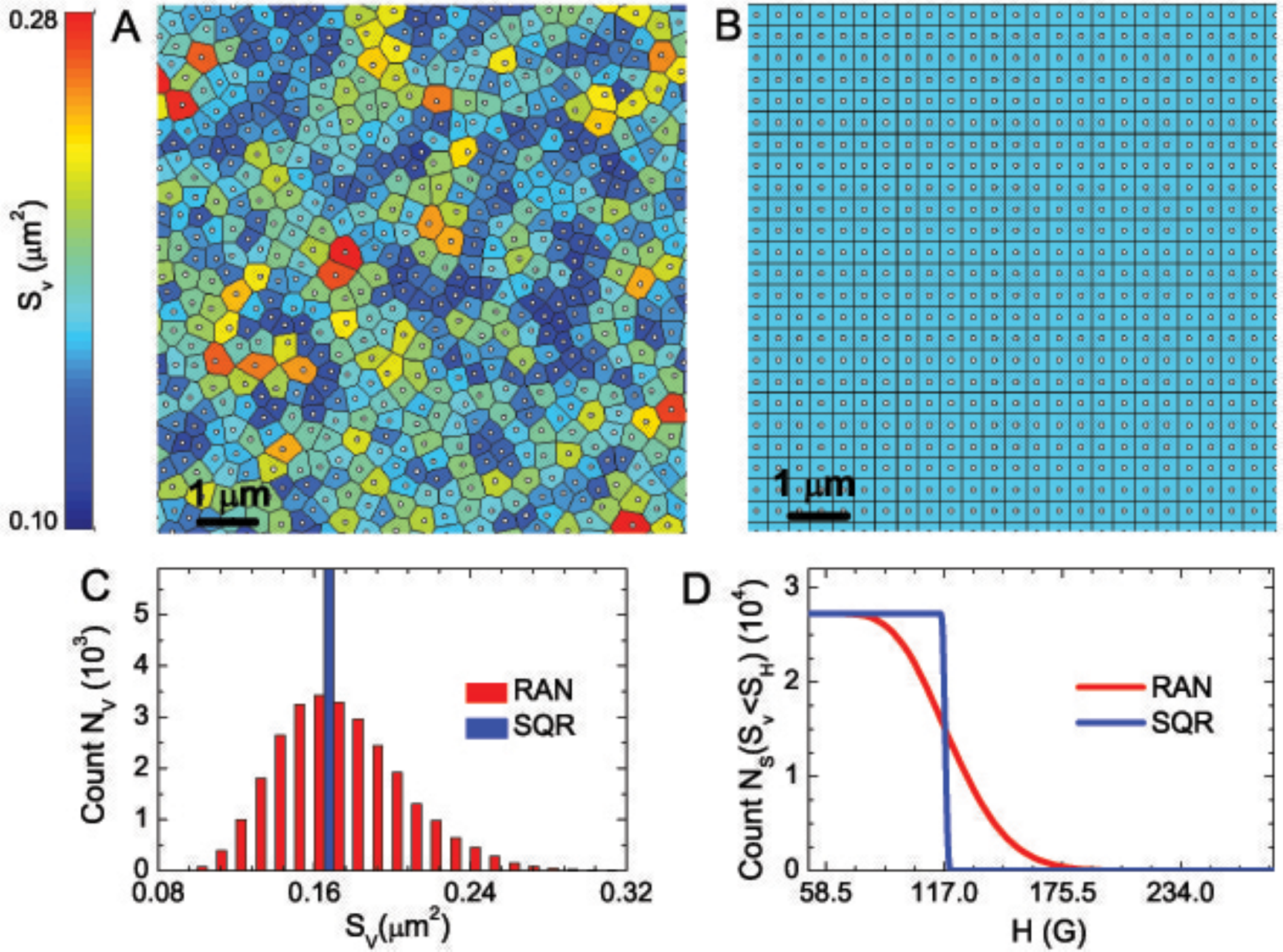}}
  \caption{\label{fig:fig3} Distributions of \textit{local density of pinning sites} (LDOPS) in Sample I. (\textbf{A}) and (\textbf{B}) Voronoi diagrams for partials of Sections RAN and SQR, respectively. The color encodes Voronoi polygon area $S_V$. (\textbf{C}) Histogram of the all Voronoi polygons. $N_V$ is the number of polygons with area $S_V$.  (\textbf{D}) Number of pinning sites corresponding to 'strong-pinning regions' as a function of magnetic field. $N_S$ is the number of polygons with polygon area $S_V<S_H$, $S_H$ is the average area occupied by each vortex at applied magnetic field $H$ and is given by $S_H=\Phi_0/H$ ($\Phi_0$ is the flux quantum).} 
  \end{center}
  \end{figure}
  
\subsection{LDOPS distribution}
 For a quantitative comparison of the LDOPS for different pinscapes we utilize the Voronoi diagram\cite{VoronoiDiagram} in which each hole is surrounded by a polygon whose sides bisect the distances to the nearest adjacent holes, as shown in Figs~\ref{fig:fig3}A and B. The area $S_V$ of the polygon of each hole is the area occupied by the hole, so the LDOPS can be represented by $1/S_V$ (the number of holes per unit area). The distribution of LDOPS can be visualized by color-coding the Voronoi polygons of different areas and quantitatively analyzed by the histogram of polygon areas, i.e. the number $N_V$ of polygons (with  area $S_V$) as a function of $S_V$. Fig.\ref{fig:fig3}A and B show the color-coded Voronoi diagrams for part of the hole-arrays in Sections RAN and SQR, respectively. Different colors in Fig.\ref{fig:fig3}A indicate the LDOPS variation in Section RAN while the uniform color in Fig.\ref{fig:fig3}b implies a uniform LDOPS in Section SQR. Fig.\ref{fig:fig3}C shows the histograms for all holes in Sections RAN and SQR. It is a vertical line for Section SQR with a single $S_V$ value of $0.176$ $\mu$m$^2$ for all Voronoi polygons, while the $S_V$ in the random section has a wide range of distribution. For a given magnetic field $H$ the number of vortices will be less than that of the holes in a local region where Voronoi polygons have $S_V\leq N_m\Phi_0/H$, where $N_m$ is the maximum number of vortices that can be trapped by each hole and $\Phi_0$ is the flux quantum. As discussed above, the sharp decrease in the critical current in Section SQR at $H_1$ indicates that $N_m=1$ for our samples\cite{ConformalWang}. In a local region with polygons having $S_V\leq S_H$ ($S_H=\Phi_0/H$), all vortices can be pinned by the holes. We refer to this as a 'strong-pinning region' for the applied magnetic field $H$. Similarly, a local region with polygons having $S_V>S_H$ will be a 'weak-pinning region' for the same field $H$ since not all vortices can be trapped in the holes, and some of them exist as interstitial vortices. In Fig.\ref{fig:fig3}D we plot the number $N_S$ of holes associated with the 'strong-pinning region' as a function of magnetic field $H$. The results clearly indicate that Section RAN has more strong-pinning regions than Section SQR when the applied magnetic field exceeds $H_1$. The $N_S(H)$ relationship around $H = 117$ G also correctly describes the $I_c(H)$ curves in Fig.\ref{fig:fig2}A for both Sections RAN and SQR. That is, the critical currents of Section RAN are lower (higher) than those of Section SQR at $H < 117$ G ($H > 117$ G) (see Fig.\ref{fig:fig3}D) and the $I_c(H)$ curve in Section SQR decays faster at around $117$ G.

The above discussion based on the number of 'strong-pinning regions', however, is unable to explain the larger critical currents in Section RAN at high magnetic fields (e.g. $H > 240$ G) where $N_S$ in both Sections RAN and SQR diminish.  Once $N_S$ becomes zero, all holes are occupied by vortices and the dissipation is determined by the motion of interstitial vortices. Due to the variation in LDOPS, i.e. non-uniform distribution of the pinning sites, the interstitial vortices in Section RAN cannot channel between the rows of holes. They may even be jammed, as proposed in Ref.\cite{Disorder2007Reichhardt} for a periodic pinscape with missing pins. In this regard, the larger critical current at high fields in Section RAN can originate from the LDOPS variation due to the suppression of the motion of interstitial vortices, in addition to broadening the magnetic field range in which 'strong-pinning regions' can exist. 

\subsection{Random, gradient and conformal}
In order to further demonstrate the importance of the spatial distribution of LDOPS on the enhancement of critical current at high magnetic fields, we fabricated another sample (Sample II), which contains sections with different distributions of LDOPS: randomly distributed holes without (Section Uni-RAN) and with (Section Grad-RAN) a global gradient. A section with a conformal array of hexagonal lattice of holes (Section Hex-CON) with the same global gradient as that in Section Grad-RAN was also patterned on the same microbridge for comparison. The number and average overall density of pinning sites are identical for all three sections. Please see Appendix for detailed method of creating the gradient pinscapes.

   \begin{figure}[tbh]
    \begin{center}
   \centerline{\includegraphics[width=0.48\textwidth]{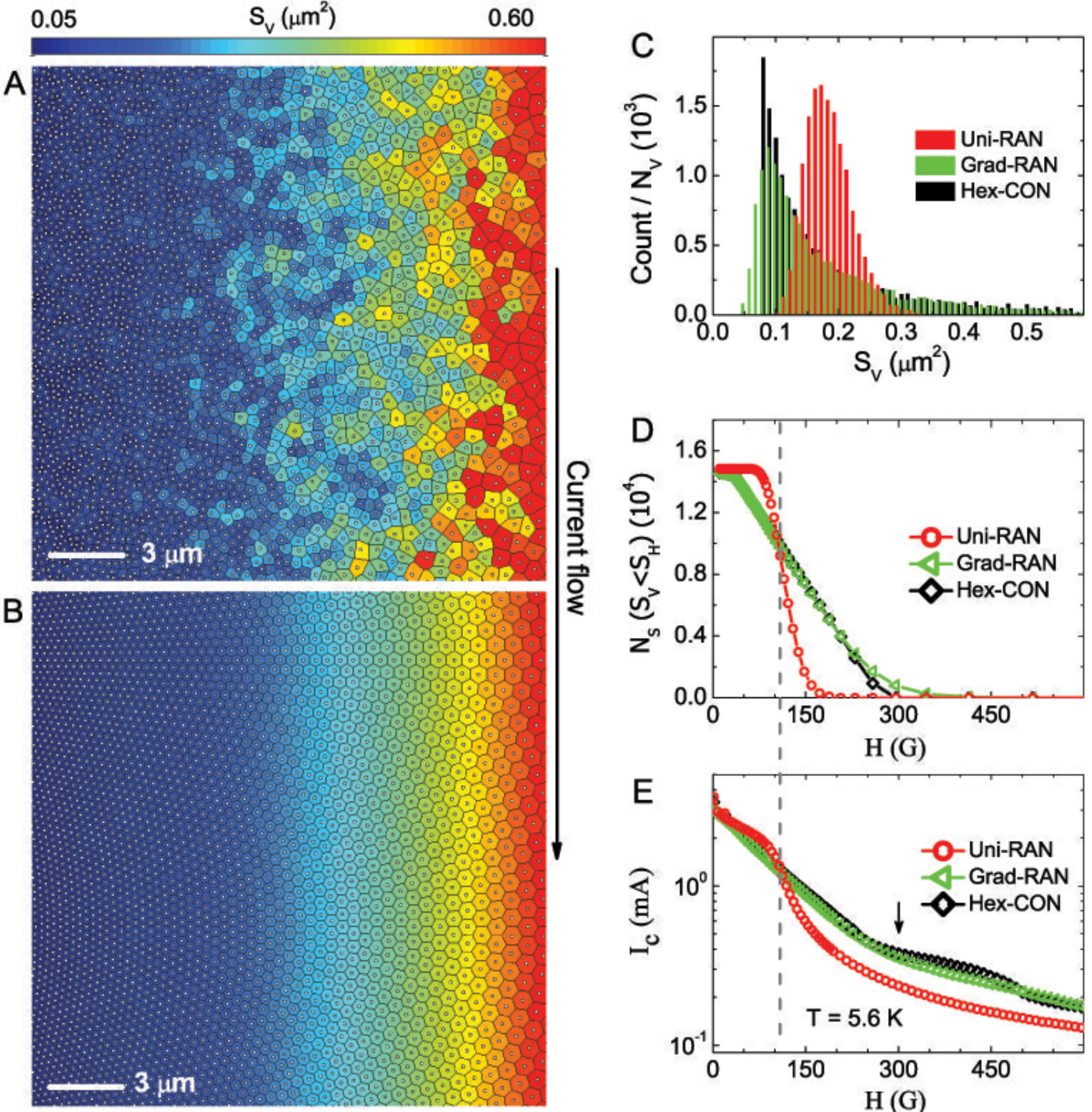}}
   \caption{\label{fig:fig4} Distributions of \textit{local density of pinning sites} (LDOPS) and critical currents of Sample II. (\textbf{A}) and (\textbf{B})Voronoi diagrams of partials of Sections Grad-RAN and Hex-CON, respectively. The current flow direction is indicated by arrow. The color presents the spatial evolution of LDOPS. (\textbf{C}) Histogram of the Voronoi polygons. (\textbf{D}) Number of pinning sites corresponding to 'strong-pinning regions' as a function of magnetic field. (\textbf{E}) Magnetic field dependence of critical currents obtained at $5.6$ K for all three sections. The curves for Sections Grad-RAN and Hex-CON in (D) and (E) nearly overlap.}
  \end{center}
  \end{figure}

We followed the same Voronoi diagram approach as in Fig.\ref{fig:fig3} to determine the LDOPS distribution in all of the three sections. Figure \ref{fig:fig4}A and \ref{fig:fig4}B show partial Voronoi diagrams for Sections Grad-RAN and Hex-CON, respectively. The color evolution of $S_V$ clearly delineates the LDOPS gradient in both sections. The histograms of the Voronoi polygon area for all three sections are shown in Fig.\ref{fig:fig4}C. They indicate that the LDOPS distribution in Section Grad-RAN and Section Hex-CON are nearly the same but much wider than that in Section Uni-RAN. Similarly, the number $N_S$ of holes in the 'strong-pinning regions' of Section Grad-RAN and Section Hex-CON have the same magnetic field dependence and fall to zero at field values higher than that for Section Uni-RAN (see Fig.\ref{fig:fig4} D). If the LDOPS distribution is the determining factor for the enhancement of the critical currents at high fields, we expect to see comparable critical currents in Section Grad-RAN and Section Hex-CON and smaller values in Section Uni-RAN. As presented in Fig.\ref{fig:fig4}E for the critical currents of all three sections measured at $T = 5.6$ K in various magnetic fields, the experimental results confirm our expectation: the $I_c(H)$ curves for Section Grad-RAN and Hex-CON, which has the same LDOPS distribution, nearly overlap with each other. In comparison to Section Uni-RAN, Section Grad-RAN and Hex-CON have a wider LDOPS distribution and thus have higher critical currents at high magnetic fields.  On the other hand, they have smaller critical currents at low magnetic fields (Fig.\ref{fig:fig4}E) due to less 'strong-pinning regions' (Fig. \ref{fig:fig4}D).

We also noticed that the $I_c$ of Section Hex-CON shows a slight up-turn at around $300$ G, as indicated by the black arrow in Fig.\ref{fig:fig4}E. This field value roughly corresponds to the first matching field for the region with the densest LDOPS. Thus, the slight improvement in $I_c$  of Section Hex-CON compared to that in Section Grad-RAN at $H > 300$ G probably originates from local vortex caging effect. That is, the local ordering in a conformal array can have detectable contributions to vortex pinning through local caging effect, while the role of local commensurate effect could not be resolved with in our experimental data. The overall enhancement of the critical current at high magnetic fields is dominated by the LDOPS distribution.

\subsection{Considerations for reliable comparisons in patterning experiments with non-uniform LDOPS}\label{Considerations}

\begin{figure}[tbh]
    \begin{center}
   \centerline{\includegraphics[width=0.45\textwidth]{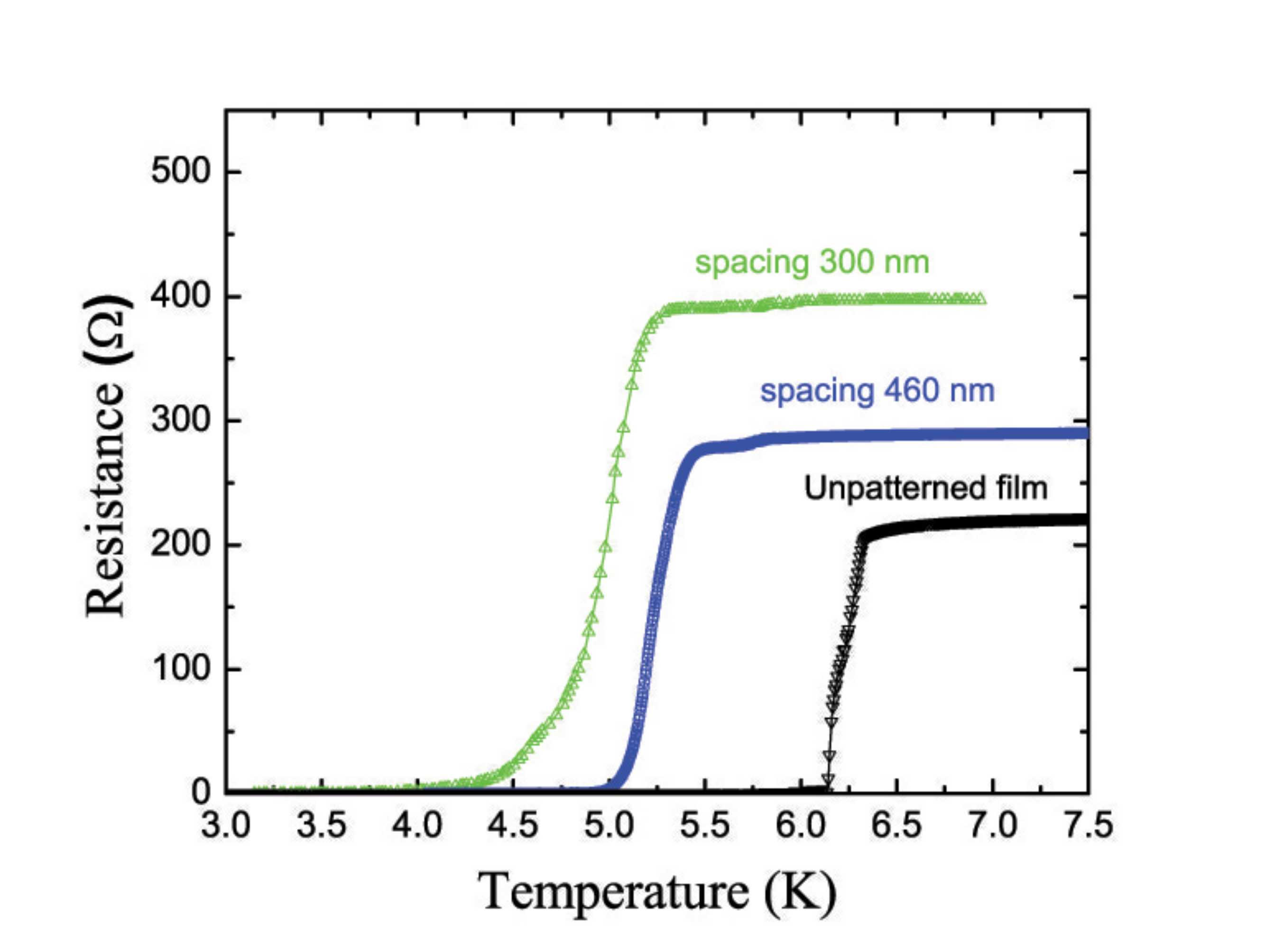}}
   \caption{\label{fig:fig5} Superconducting transition curves of focused-ion-beam (FIB) patterned MoGe sections containing triangular arrays of holes with two different hole-hole spacings ($460$ nm and $300$ nm) and a unpatterned reference section. In contrast to the sample patterning method using E-beam lithography followed by reactive ion etching (see Fig.~\ref{fig:fig1}F), the $T_c$ of FIB patterned sample is suppressed in the two patterned sections and decreases with shorter hole-hole spacing.}
  \end{center}
  \end{figure}

Introducing artificial pinning sites in a superconducting thin film does not only enhance vortex pinning but also could damage the sample. The $T_c$ of the sample is usually suppressed due to processing-induced sample degradation. Figure~\ref{fig:fig5} shows the R-T curves of a reference MoGe film and two focused-ion-beam (FIB) patterned MoGe films containing triangular array of holes with different hole-hole spacings. The $T_c$s of the two patterned films are lower than the unpatterned film, indicating processing-induced degradation of superconductivity. Furthermore, the film with hole-hole spacing of $300$ nm has lower $T_c$ than that of $460$ nm one. That is, when holes are patterned with a process that can degrade the sample, the higher the hole density, the lower the $T_c$. Such effects are frequently observed in FIB patterned samples and samples fabricated using direct lift-off method.

When using patterning process that degrades superconductivity as described above, the samples with different pinning geometries can have different $T_c$s, which confounds the comparison of different pinning landscapes.  Due to the variation in $T_c$, one usually compares the results at the same reduced temperature $T/T_c$. However for a non-uniform pinning geometry, the $T_c$ of the sample will be spatially inhomogeneous because the nearest hole-hole spacings are spatially different. That is, the region with higher LDOPS will have lower $T_c$. The $T_c$ determined from the transport measurements gives values from the current channels with highest $T_c$, resulting in an over-estimate of the $T_c$. Thus, when studying the randomly distributed pinning geometry of holes, it is important to adopt a sample fabrication method that limits the damage only to the hole area and not influence the $T_c$ of the rest of the sample.

Our sample patterning method of using E-beam lithography followed by reactive ion etching does not suppress the $T_c$ of the films (Fig.~\ref{fig:fig1}F). The film is protected by an e-beam resist mask during the etching process and there is no damage to the film except for the area of the hole, which not only enables us to measure and compare the results at the same temperature but also produce holes with strong pinning potential.

\section{Conclusion}
In summary, we have demonstrated the pinning effects of random hole-arrays with and without a global gradient by comparing them with those of a square-array and a conformal array of a hexagonal lattice of holes. We used a Voronoi diagram approach to visualize and quantify the distribution of \textit{local density of pinning sites} (LDOPS), revealing the key role of LDOPS distribution in enhancing the critical current at high magnetic fields. We found that a properly tailored LDOPS in a random orientation has the best potential for enhancing the critical current over a wide range of magnetic fields. Our results provide a deeper insight to the often overlooked random pinscape, and should foster its study in other related systems,  such as colloidal crystals,\cite{Colloidal,Pertsinidis2001,Gokhale2014} Bose-Einstein condensates\cite{BES} and Luttinger liquid of strongly interacting bosons.\cite{Haller2010LuttingerLiquid}

\begin{acknowledgments}
This work was supported by the U.S. Department of Energy, Office of Science, Basic Energy Sciences, Materials Sciences and Engineering Division. Nano patterning and morphological analysis were performed at Argonne’s Center for Nanoscale Materials (CNM) and the Electron Microscopy Center (EMC), which are supported by DOE, Office of Science, BES. L.R.T. and Z.L.X. acknowledge NSF Grant No.
DMR-1407175 and NIU’s Nanoscience Graduate Fellowship.
\end{acknowledgments}

\appendix*
\section{Creation of gradient pinscapes using conformal transformation}
 The coordinates of gradient pinscapes were generated from conformal transformation of a uniform random array and a hexagonal array. The average hole density of the uniform random array and that of the hexagonal array are identical, so that the transformed gradient arrays have the same global gradient. We use conformal transformation\cite{ConformalRayPRL,ConformalWang} of site position ($x$, $y$)s of a uniform pinning array to generate pinning sites position ($x'$, $y'$)s of the gradient pinscapes.  A lower limit of $200$ nm was set for the center to center of hole separation for the random array used to create the random gradient array in Sample II. First, the pinning site coordinates ($x$,$y$)s in an area of $r_{in}<\sqrt{x^2+y^2}<r_{out}$ were created. Then ($x$, $y$) were converted to ($x'$, $y'$) using the following formula:
\[ \begin{cases}
x'=\begin{cases}
r_{out}[\arctan(\frac{y}{x})+\pi] & \quad \text{if } y\leq 0\\
r_{out}\arctan(\frac{y}{x}) & \quad \text{if } y>0
\end{cases}\\
y'=\frac{1}{2}r_{out}\ln(\frac{r_{out}^2}{x^2+y^2})
\end{cases}\]
In Sample II we use $r_{in} = 7.9$ $\mu$m and $r_{out} = 22.5$ $\mu$m. Two generated conformal arrays are placed to face each other (similar to that in Ref \cite{ConformalWang}) with a separation of $0.5$ $\mu$m which is roughly the distance between holes in the center region. The total number of holes in the conformal transformed array is about $15000$. The corresponding uniform-random array is created with the same number of holes and covering an area with the same size.

\bibliography{Random}

\begin{thebibliography}{60}%
\makeatletter
\providecommand \@ifxundefined [1]{%
 \@ifx{#1\undefined}
}%
\providecommand \@ifnum [1]{%
 \ifnum #1\expandafter \@firstoftwo
 \else \expandafter \@secondoftwo
 \fi
}%
\providecommand \@ifx [1]{%
 \ifx #1\expandafter \@firstoftwo
 \else \expandafter \@secondoftwo
 \fi
}%
\providecommand \natexlab [1]{#1}%
\providecommand \enquote  [1]{``#1''}%
\providecommand \bibnamefont  [1]{#1}%
\providecommand \bibfnamefont [1]{#1}%
\providecommand \citenamefont [1]{#1}%
\providecommand \href@noop [0]{\@secondoftwo}%
\providecommand \href [0]{\begingroup \@sanitize@url \@href}%
\providecommand \@href[1]{\@@startlink{#1}\@@href}%
\providecommand \@@href[1]{\endgroup#1\@@endlink}%
\providecommand \@sanitize@url [0]{\catcode `\\12\catcode `\$12\catcode
  `\&12\catcode `\#12\catcode `\^12\catcode `\_12\catcode `\%12\relax}%
\providecommand \@@startlink[1]{}%
\providecommand \@@endlink[0]{}%
\providecommand \url  [0]{\begingroup\@sanitize@url \@url }%
\providecommand \@url [1]{\endgroup\@href {#1}{\urlprefix }}%
\providecommand \urlprefix  [0]{URL }%
\providecommand \Eprint [0]{\href }%
\providecommand \doibase [0]{http://dx.doi.org/}%
\providecommand \selectlanguage [0]{\@gobble}%
\providecommand \bibinfo  [0]{\@secondoftwo}%
\providecommand \bibfield  [0]{\@secondoftwo}%
\providecommand \translation [1]{[#1]}%
\providecommand \BibitemOpen [0]{}%
\providecommand \bibitemStop [0]{}%
\providecommand \bibitemNoStop [0]{.\EOS\space}%
\providecommand \EOS [0]{\spacefactor3000\relax}%
\providecommand \BibitemShut  [1]{\csname bibitem#1\endcsname}%
\let\auto@bib@innerbib\@empty
\bibitem [{\citenamefont {Chong}\ \emph {et~al.}(1997)\citenamefont {Chong},
  \citenamefont {Hiroi}, \citenamefont {Izumi}, \citenamefont {Shimoyama},
  \citenamefont {Nakayama}, \citenamefont {Kishio}, \citenamefont {Terashima},
  \citenamefont {Bando},\ and\ \citenamefont {Takano}}]{Ref1}%
  \BibitemOpen
  \bibfield  {author} {\bibinfo {author} {\bibfnamefont {I.}~\bibnamefont
  {Chong}}, \bibinfo {author} {\bibfnamefont {Z.}~\bibnamefont {Hiroi}},
  \bibinfo {author} {\bibfnamefont {M.}~\bibnamefont {Izumi}}, \bibinfo
  {author} {\bibfnamefont {J.}~\bibnamefont {Shimoyama}}, \bibinfo {author}
  {\bibfnamefont {Y.}~\bibnamefont {Nakayama}}, \bibinfo {author}
  {\bibfnamefont {K.}~\bibnamefont {Kishio}}, \bibinfo {author} {\bibfnamefont
  {T.}~\bibnamefont {Terashima}}, \bibinfo {author} {\bibfnamefont
  {Y.}~\bibnamefont {Bando}}, \ and\ \bibinfo {author} {\bibfnamefont
  {M.}~\bibnamefont {Takano}},\ }\href {\doibase 10.1126/science.276.5313.770}
  {\bibfield  {journal} {\bibinfo  {journal} {Science}\ }\textbf {\bibinfo
  {volume} {276}},\ \bibinfo {pages} {770} (\bibinfo {year}
  {1997})}\BibitemShut {NoStop}%
\bibitem [{\citenamefont {Crabtree}\ and\ \citenamefont {Nelson}(1997)}]{Ref2}%
  \BibitemOpen
  \bibfield  {author} {\bibinfo {author} {\bibfnamefont {G.~W.}\ \bibnamefont
  {Crabtree}}\ and\ \bibinfo {author} {\bibfnamefont {D.~R.}\ \bibnamefont
  {Nelson}},\ }\href {\doibase 10.1063/1.881715} {\bibfield  {journal}
  {\bibinfo  {journal} {Physics Today}\ }\textbf {\bibinfo {volume} {50}},\
  \bibinfo {pages} {38} (\bibinfo {year} {1997})}\BibitemShut {NoStop}%
\bibitem [{\citenamefont {Martin}\ \emph {et~al.}(1997)\citenamefont {Martin},
  \citenamefont {Velez}, \citenamefont {Nogues},\ and\ \citenamefont
  {Schuller}}]{Ref3}%
  \BibitemOpen
  \bibfield  {author} {\bibinfo {author} {\bibfnamefont {J.~I.}\ \bibnamefont
  {Martin}}, \bibinfo {author} {\bibfnamefont {M.}~\bibnamefont {Velez}},
  \bibinfo {author} {\bibfnamefont {J.}~\bibnamefont {Nogues}}, \ and\ \bibinfo
  {author} {\bibfnamefont {I.~K.}\ \bibnamefont {Schuller}},\ }\href {\doibase
  10.1103/PhysRevLett.79.1929} {\bibfield  {journal} {\bibinfo  {journal}
  {Phys. Rev. Lett.}\ }\textbf {\bibinfo {volume} {79}},\ \bibinfo {pages}
  {1929} (\bibinfo {year} {1997})}\BibitemShut {NoStop}%
\bibitem [{\citenamefont {Haugan}\ \emph {et~al.}(2004)\citenamefont {Haugan},
  \citenamefont {Barnes}, \citenamefont {Wheeler}, \citenamefont
  {Meisenkothen},\ and\ \citenamefont {Sumption}}]{Ref4}%
  \BibitemOpen
  \bibfield  {author} {\bibinfo {author} {\bibfnamefont {T.}~\bibnamefont
  {Haugan}}, \bibinfo {author} {\bibfnamefont {P.~N.}\ \bibnamefont {Barnes}},
  \bibinfo {author} {\bibfnamefont {R.}~\bibnamefont {Wheeler}}, \bibinfo
  {author} {\bibfnamefont {F.}~\bibnamefont {Meisenkothen}}, \ and\ \bibinfo
  {author} {\bibfnamefont {M.}~\bibnamefont {Sumption}},\ }\href@noop {}
  {\bibfield  {journal} {\bibinfo  {journal} {Nature}\ }\textbf {\bibinfo
  {volume} {430}},\ \bibinfo {pages} {867} (\bibinfo {year}
  {2004})}\BibitemShut {NoStop}%
\bibitem [{\citenamefont {Dam}\ \emph {et~al.}(1999)\citenamefont {Dam},
  \citenamefont {Huijbregtse}, \citenamefont {Klaassen}, \citenamefont {van~der
  Geest}, \citenamefont {Doornbos}, \citenamefont {Rector}, \citenamefont
  {Testa}, \citenamefont {Freisem}, \citenamefont {Martinez}, \citenamefont
  {Stauble-Pumpin},\ and\ \citenamefont {Griessen}}]{Dam1999Ref5}%
  \BibitemOpen
  \bibfield  {author} {\bibinfo {author} {\bibfnamefont {B.}~\bibnamefont
  {Dam}}, \bibinfo {author} {\bibfnamefont {J.~M.}\ \bibnamefont
  {Huijbregtse}}, \bibinfo {author} {\bibfnamefont {F.~C.}\ \bibnamefont
  {Klaassen}}, \bibinfo {author} {\bibfnamefont {R.~C.~F.}\ \bibnamefont
  {van~der Geest}}, \bibinfo {author} {\bibfnamefont {G.}~\bibnamefont
  {Doornbos}}, \bibinfo {author} {\bibfnamefont {J.~H.}\ \bibnamefont
  {Rector}}, \bibinfo {author} {\bibfnamefont {A.~M.}\ \bibnamefont {Testa}},
  \bibinfo {author} {\bibfnamefont {S.}~\bibnamefont {Freisem}}, \bibinfo
  {author} {\bibfnamefont {J.~C.}\ \bibnamefont {Martinez}}, \bibinfo {author}
  {\bibfnamefont {B.}~\bibnamefont {Stauble-Pumpin}}, \ and\ \bibinfo {author}
  {\bibfnamefont {R.}~\bibnamefont {Griessen}},\ }\href@noop {} {\bibfield
  {journal} {\bibinfo  {journal} {Nature}\ }\textbf {\bibinfo {volume} {399}},\
  \bibinfo {pages} {439} (\bibinfo {year} {1999})}\BibitemShut {NoStop}%
\bibitem [{\citenamefont {Bugoslavsky}\ \emph {et~al.}(2001)\citenamefont
  {Bugoslavsky}, \citenamefont {Cohen}, \citenamefont {Perkins}, \citenamefont
  {Polichetti}, \citenamefont {Tate}, \citenamefont {Gwilliam},\ and\
  \citenamefont {Caplin}}]{Bugoslavsky2001Ref6}%
  \BibitemOpen
  \bibfield  {author} {\bibinfo {author} {\bibfnamefont {Y.}~\bibnamefont
  {Bugoslavsky}}, \bibinfo {author} {\bibfnamefont {L.~F.}\ \bibnamefont
  {Cohen}}, \bibinfo {author} {\bibfnamefont {G.~K.}\ \bibnamefont {Perkins}},
  \bibinfo {author} {\bibfnamefont {M.}~\bibnamefont {Polichetti}}, \bibinfo
  {author} {\bibfnamefont {T.~J.}\ \bibnamefont {Tate}}, \bibinfo {author}
  {\bibfnamefont {R.}~\bibnamefont {Gwilliam}}, \ and\ \bibinfo {author}
  {\bibfnamefont {A.~D.}\ \bibnamefont {Caplin}},\ }\href@noop {} {\bibfield
  {journal} {\bibinfo  {journal} {Nature}\ }\textbf {\bibinfo {volume} {411}},\
  \bibinfo {pages} {561} (\bibinfo {year} {2001})}\BibitemShut {NoStop}%
\bibitem [{\citenamefont {Kang}\ \emph {et~al.}(2006)\citenamefont {Kang},
  \citenamefont {Goyal}, \citenamefont {Li}, \citenamefont {Gapud},
  \citenamefont {Martin}, \citenamefont {Heatherly}, \citenamefont {Thompson},
  \citenamefont {Christen}, \citenamefont {List}, \citenamefont {Paranthaman},\
  and\ \citenamefont {Lee}}]{Ref5}%
  \BibitemOpen
  \bibfield  {author} {\bibinfo {author} {\bibfnamefont {S.}~\bibnamefont
  {Kang}}, \bibinfo {author} {\bibfnamefont {A.}~\bibnamefont {Goyal}},
  \bibinfo {author} {\bibfnamefont {J.}~\bibnamefont {Li}}, \bibinfo {author}
  {\bibfnamefont {A.~A.}\ \bibnamefont {Gapud}}, \bibinfo {author}
  {\bibfnamefont {P.~M.}\ \bibnamefont {Martin}}, \bibinfo {author}
  {\bibfnamefont {L.}~\bibnamefont {Heatherly}}, \bibinfo {author}
  {\bibfnamefont {J.~R.}\ \bibnamefont {Thompson}}, \bibinfo {author}
  {\bibfnamefont {D.~K.}\ \bibnamefont {Christen}}, \bibinfo {author}
  {\bibfnamefont {F.~A.}\ \bibnamefont {List}}, \bibinfo {author}
  {\bibfnamefont {M.}~\bibnamefont {Paranthaman}}, \ and\ \bibinfo {author}
  {\bibfnamefont {D.~F.}\ \bibnamefont {Lee}},\ }\href {\doibase
  10.1126/science.1124872} {\bibfield  {journal} {\bibinfo  {journal}
  {Science}\ }\textbf {\bibinfo {volume} {311}},\ \bibinfo {pages} {1911}
  (\bibinfo {year} {2006})}\BibitemShut {NoStop}%
\bibitem [{\citenamefont {Llordes}\ \emph {et~al.}(2012)\citenamefont
  {Llordes}, \citenamefont {Palau}, \citenamefont {Gazquez}, \citenamefont
  {Coll}, \citenamefont {Vlad}, \citenamefont {Pomar}, \citenamefont {Arbiol},
  \citenamefont {Guzman}, \citenamefont {Ye}, \citenamefont {Rouco},
  \citenamefont {Sandiumenge}, \citenamefont {Ricart}, \citenamefont {Puig},
  \citenamefont {Varela}, \citenamefont {Chateigner}, \citenamefont {Vanacken},
  \citenamefont {Gutierrez}, \citenamefont {Moshchalkov}, \citenamefont
  {Deutscher}, \citenamefont {Magen},\ and\ \citenamefont {Obradors}}]{Ref6}%
  \BibitemOpen
  \bibfield  {author} {\bibinfo {author} {\bibfnamefont {A.}~\bibnamefont
  {Llordes}}, \bibinfo {author} {\bibfnamefont {A.}~\bibnamefont {Palau}},
  \bibinfo {author} {\bibfnamefont {J.}~\bibnamefont {Gazquez}}, \bibinfo
  {author} {\bibfnamefont {M.}~\bibnamefont {Coll}}, \bibinfo {author}
  {\bibfnamefont {R.}~\bibnamefont {Vlad}}, \bibinfo {author} {\bibfnamefont
  {A.}~\bibnamefont {Pomar}}, \bibinfo {author} {\bibfnamefont
  {J.}~\bibnamefont {Arbiol}}, \bibinfo {author} {\bibfnamefont
  {R.}~\bibnamefont {Guzman}}, \bibinfo {author} {\bibfnamefont
  {S.}~\bibnamefont {Ye}}, \bibinfo {author} {\bibfnamefont {V.}~\bibnamefont
  {Rouco}}, \bibinfo {author} {\bibfnamefont {F.}~\bibnamefont {Sandiumenge}},
  \bibinfo {author} {\bibfnamefont {S.}~\bibnamefont {Ricart}}, \bibinfo
  {author} {\bibfnamefont {T.}~\bibnamefont {Puig}}, \bibinfo {author}
  {\bibfnamefont {M.}~\bibnamefont {Varela}}, \bibinfo {author} {\bibfnamefont
  {D.}~\bibnamefont {Chateigner}}, \bibinfo {author} {\bibfnamefont
  {J.}~\bibnamefont {Vanacken}}, \bibinfo {author} {\bibfnamefont
  {J.}~\bibnamefont {Gutierrez}}, \bibinfo {author} {\bibfnamefont
  {V.}~\bibnamefont {Moshchalkov}}, \bibinfo {author} {\bibfnamefont
  {G.}~\bibnamefont {Deutscher}}, \bibinfo {author} {\bibfnamefont
  {C.}~\bibnamefont {Magen}}, \ and\ \bibinfo {author} {\bibfnamefont
  {X.}~\bibnamefont {Obradors}},\ }\href@noop {} {\bibfield  {journal}
  {\bibinfo  {journal} {Nat. Mater.}\ }\textbf {\bibinfo {volume} {11}},\
  \bibinfo {pages} {329} (\bibinfo {year} {2012})}\BibitemShut {NoStop}%
\bibitem [{\citenamefont {Moshchalkov}\ \emph {et~al.}(1998)\citenamefont
  {Moshchalkov}, \citenamefont {Baert}, \citenamefont {Metlushko},
  \citenamefont {Rosseel}, \citenamefont {Van~Bael}, \citenamefont {Temst},
  \citenamefont {Bruynseraede},\ and\ \citenamefont {Jonckheere}}]{Ref10}%
  \BibitemOpen
  \bibfield  {author} {\bibinfo {author} {\bibfnamefont {V.~V.}\ \bibnamefont
  {Moshchalkov}}, \bibinfo {author} {\bibfnamefont {M.}~\bibnamefont {Baert}},
  \bibinfo {author} {\bibfnamefont {V.~V.}\ \bibnamefont {Metlushko}}, \bibinfo
  {author} {\bibfnamefont {E.}~\bibnamefont {Rosseel}}, \bibinfo {author}
  {\bibfnamefont {M.~J.}\ \bibnamefont {Van~Bael}}, \bibinfo {author}
  {\bibfnamefont {K.}~\bibnamefont {Temst}}, \bibinfo {author} {\bibfnamefont
  {Y.}~\bibnamefont {Bruynseraede}}, \ and\ \bibinfo {author} {\bibfnamefont
  {R.}~\bibnamefont {Jonckheere}},\ }\href {\doibase 10.1103/PhysRevB.57.3615}
  {\bibfield  {journal} {\bibinfo  {journal} {Phys. Rev. B}\ }\textbf {\bibinfo
  {volume} {57}},\ \bibinfo {pages} {3615} (\bibinfo {year}
  {1998})}\BibitemShut {NoStop}%
\bibitem [{\citenamefont {Kemmler}\ \emph {et~al.}(2006)\citenamefont
  {Kemmler}, \citenamefont {G\"urlich}, \citenamefont {Sterck}, \citenamefont
  {P\"ohler}, \citenamefont {Neuhaus}, \citenamefont {Siegel}, \citenamefont
  {Kleiner},\ and\ \citenamefont {Koelle}}]{Ref14}%
  \BibitemOpen
  \bibfield  {author} {\bibinfo {author} {\bibfnamefont {M.}~\bibnamefont
  {Kemmler}}, \bibinfo {author} {\bibfnamefont {C.}~\bibnamefont {G\"urlich}},
  \bibinfo {author} {\bibfnamefont {A.}~\bibnamefont {Sterck}}, \bibinfo
  {author} {\bibfnamefont {H.}~\bibnamefont {P\"ohler}}, \bibinfo {author}
  {\bibfnamefont {M.}~\bibnamefont {Neuhaus}}, \bibinfo {author} {\bibfnamefont
  {M.}~\bibnamefont {Siegel}}, \bibinfo {author} {\bibfnamefont
  {R.}~\bibnamefont {Kleiner}}, \ and\ \bibinfo {author} {\bibfnamefont
  {D.}~\bibnamefont {Koelle}},\ }\href {\doibase 10.1103/PhysRevLett.97.147003}
  {\bibfield  {journal} {\bibinfo  {journal} {Phys. Rev. Lett.}\ }\textbf
  {\bibinfo {volume} {97}},\ \bibinfo {pages} {147003} (\bibinfo {year}
  {2006})}\BibitemShut {NoStop}%
\bibitem [{\citenamefont {Velez}\ \emph {et~al.}(2008)\citenamefont {Velez},
  \citenamefont {Martin}, \citenamefont {Villegas}, \citenamefont {Hoffmann},
  \citenamefont {Gonzalez}, \citenamefont {Vicent},\ and\ \citenamefont
  {Schuller}}]{Ref9}%
  \BibitemOpen
  \bibfield  {author} {\bibinfo {author} {\bibfnamefont {M.}~\bibnamefont
  {Velez}}, \bibinfo {author} {\bibfnamefont {J.}~\bibnamefont {Martin}},
  \bibinfo {author} {\bibfnamefont {J.}~\bibnamefont {Villegas}}, \bibinfo
  {author} {\bibfnamefont {A.}~\bibnamefont {Hoffmann}}, \bibinfo {author}
  {\bibfnamefont {E.}~\bibnamefont {Gonzalez}}, \bibinfo {author}
  {\bibfnamefont {J.}~\bibnamefont {Vicent}}, \ and\ \bibinfo {author}
  {\bibfnamefont {I.~K.}\ \bibnamefont {Schuller}},\ }\href@noop {} {\bibfield
  {journal} {\bibinfo  {journal} {J. Magn. Magn. Mater.}\ }\textbf {\bibinfo
  {volume} {320}},\ \bibinfo {pages} {2547} (\bibinfo {year}
  {2008})}\BibitemShut {NoStop}%
\bibitem [{\citenamefont {Welp}\ \emph {et~al.}(2002)\citenamefont {Welp},
  \citenamefont {Xiao}, \citenamefont {Jiang}, \citenamefont {Vlasko-Vlasov},
  \citenamefont {Bader}, \citenamefont {Crabtree}, \citenamefont {Liang},
  \citenamefont {Chik},\ and\ \citenamefont {Xu}}]{Ref11}%
  \BibitemOpen
  \bibfield  {author} {\bibinfo {author} {\bibfnamefont {U.}~\bibnamefont
  {Welp}}, \bibinfo {author} {\bibfnamefont {Z.~L.}\ \bibnamefont {Xiao}},
  \bibinfo {author} {\bibfnamefont {J.~S.}\ \bibnamefont {Jiang}}, \bibinfo
  {author} {\bibfnamefont {V.~K.}\ \bibnamefont {Vlasko-Vlasov}}, \bibinfo
  {author} {\bibfnamefont {S.~D.}\ \bibnamefont {Bader}}, \bibinfo {author}
  {\bibfnamefont {G.~W.}\ \bibnamefont {Crabtree}}, \bibinfo {author}
  {\bibfnamefont {J.}~\bibnamefont {Liang}}, \bibinfo {author} {\bibfnamefont
  {H.}~\bibnamefont {Chik}}, \ and\ \bibinfo {author} {\bibfnamefont {J.~M.}\
  \bibnamefont {Xu}},\ }\href {\doibase 10.1103/PhysRevB.66.212507} {\bibfield
  {journal} {\bibinfo  {journal} {Phys. Rev. B}\ }\textbf {\bibinfo {volume}
  {66}},\ \bibinfo {pages} {212507} (\bibinfo {year} {2002})}\BibitemShut
  {NoStop}%
\bibitem [{\citenamefont {Silhanek}\ \emph {et~al.}(2005)\citenamefont
  {Silhanek}, \citenamefont {Van~Look}, \citenamefont {Jonckheere},
  \citenamefont {Zhu}, \citenamefont {Raedts},\ and\ \citenamefont
  {Moshchalkov}}]{CompositeAntidots}%
  \BibitemOpen
  \bibfield  {author} {\bibinfo {author} {\bibfnamefont {A.~V.}\ \bibnamefont
  {Silhanek}}, \bibinfo {author} {\bibfnamefont {L.}~\bibnamefont {Van~Look}},
  \bibinfo {author} {\bibfnamefont {R.}~\bibnamefont {Jonckheere}}, \bibinfo
  {author} {\bibfnamefont {B.~Y.}\ \bibnamefont {Zhu}}, \bibinfo {author}
  {\bibfnamefont {S.}~\bibnamefont {Raedts}}, \ and\ \bibinfo {author}
  {\bibfnamefont {V.~V.}\ \bibnamefont {Moshchalkov}},\ }\href {\doibase
  10.1103/PhysRevB.72.014507} {\bibfield  {journal} {\bibinfo  {journal} {Phys.
  Rev. B}\ }\textbf {\bibinfo {volume} {72}},\ \bibinfo {pages} {014507}
  (\bibinfo {year} {2005})}\BibitemShut {NoStop}%
\bibitem [{\citenamefont {Avci}\ \emph {et~al.}(2010)\citenamefont {Avci},
  \citenamefont {Xiao}, \citenamefont {Hua}, \citenamefont {Imre},
  \citenamefont {Divan}, \citenamefont {Pearson}, \citenamefont {Welp},
  \citenamefont {Kwok},\ and\ \citenamefont {Crabtree}}]{Ref17}%
  \BibitemOpen
  \bibfield  {author} {\bibinfo {author} {\bibfnamefont {S.}~\bibnamefont
  {Avci}}, \bibinfo {author} {\bibfnamefont {Z.~L.}\ \bibnamefont {Xiao}},
  \bibinfo {author} {\bibfnamefont {J.}~\bibnamefont {Hua}}, \bibinfo {author}
  {\bibfnamefont {A.}~\bibnamefont {Imre}}, \bibinfo {author} {\bibfnamefont
  {R.}~\bibnamefont {Divan}}, \bibinfo {author} {\bibfnamefont
  {J.}~\bibnamefont {Pearson}}, \bibinfo {author} {\bibfnamefont
  {U.}~\bibnamefont {Welp}}, \bibinfo {author} {\bibfnamefont {W.~K.}\
  \bibnamefont {Kwok}}, \ and\ \bibinfo {author} {\bibfnamefont {G.~W.}\
  \bibnamefont {Crabtree}},\ }\href {\doibase 10.1063/1.3473783} {\bibfield
  {journal} {\bibinfo  {journal} {Appl. Phys. Lett.}\ }\textbf {\bibinfo
  {volume} {97}},\ \bibinfo {eid} {042511} (\bibinfo {year}
  {2010})}\BibitemShut {NoStop}%
\bibitem [{\citenamefont {Chiliotte}\ \emph {et~al.}(2011)\citenamefont
  {Chiliotte}, \citenamefont {Pasquini}, \citenamefont {Bekeris}, \citenamefont
  {Villegas}, \citenamefont {Li},\ and\ \citenamefont {Schuller}}]{Ref18}%
  \BibitemOpen
  \bibfield  {author} {\bibinfo {author} {\bibfnamefont {C.}~\bibnamefont
  {Chiliotte}}, \bibinfo {author} {\bibfnamefont {G.}~\bibnamefont {Pasquini}},
  \bibinfo {author} {\bibfnamefont {V.}~\bibnamefont {Bekeris}}, \bibinfo
  {author} {\bibfnamefont {J.~E.}\ \bibnamefont {Villegas}}, \bibinfo {author}
  {\bibfnamefont {C.-P.}\ \bibnamefont {Li}}, \ and\ \bibinfo {author}
  {\bibfnamefont {I.~K.}\ \bibnamefont {Schuller}},\ }\href@noop {} {\bibfield
  {journal} {\bibinfo  {journal} {Supercond. Sci. Technol.}\ }\textbf {\bibinfo
  {volume} {24}},\ \bibinfo {pages} {065008} (\bibinfo {year}
  {2011})}\BibitemShut {NoStop}%
\bibitem [{\citenamefont {Wu}\ \emph {et~al.}(2007)\citenamefont {Wu},
  \citenamefont {Horng}, \citenamefont {Wu}, \citenamefont {Cao}, \citenamefont
  {Kol\'{a}\v{c}ek},\ and\ \citenamefont {Yang}}]{Ref25}%
  \BibitemOpen
  \bibfield  {author} {\bibinfo {author} {\bibfnamefont {T.~C.}\ \bibnamefont
  {Wu}}, \bibinfo {author} {\bibfnamefont {L.}~\bibnamefont {Horng}}, \bibinfo
  {author} {\bibfnamefont {J.~C.}\ \bibnamefont {Wu}}, \bibinfo {author}
  {\bibfnamefont {R.}~\bibnamefont {Cao}}, \bibinfo {author} {\bibfnamefont
  {J.}~\bibnamefont {Kol\'{a}\v{c}ek}}, \ and\ \bibinfo {author} {\bibfnamefont
  {T.~J.}\ \bibnamefont {Yang}},\ }\href {\doibase 10.1063/1.2767386}
  {\bibfield  {journal} {\bibinfo  {journal} {J. Appl. Phys.}\ }\textbf
  {\bibinfo {volume} {102}},\ \bibinfo {eid} {033918} (\bibinfo {year}
  {2007})}\BibitemShut {NoStop}%
\bibitem [{\citenamefont {Motta}\ \emph {et~al.}(2013)\citenamefont {Motta},
  \citenamefont {Colauto}, \citenamefont {Ortiz}, \citenamefont {Fritzsche},
  \citenamefont {Cuppens}, \citenamefont {Gillijns}, \citenamefont
  {Moshchalkov}, \citenamefont {Johansen}, \citenamefont {Sanchez},\ and\
  \citenamefont {Silhanek}}]{Ref26}%
  \BibitemOpen
  \bibfield  {author} {\bibinfo {author} {\bibfnamefont {M.}~\bibnamefont
  {Motta}}, \bibinfo {author} {\bibfnamefont {F.}~\bibnamefont {Colauto}},
  \bibinfo {author} {\bibfnamefont {W.~A.}\ \bibnamefont {Ortiz}}, \bibinfo
  {author} {\bibfnamefont {J.}~\bibnamefont {Fritzsche}}, \bibinfo {author}
  {\bibfnamefont {J.}~\bibnamefont {Cuppens}}, \bibinfo {author} {\bibfnamefont
  {W.}~\bibnamefont {Gillijns}}, \bibinfo {author} {\bibfnamefont {V.~V.}\
  \bibnamefont {Moshchalkov}}, \bibinfo {author} {\bibfnamefont {T.~H.}\
  \bibnamefont {Johansen}}, \bibinfo {author} {\bibfnamefont {A.}~\bibnamefont
  {Sanchez}}, \ and\ \bibinfo {author} {\bibfnamefont {A.~V.}\ \bibnamefont
  {Silhanek}},\ }\href {\doibase http://dx.doi.org/10.1063/1.4807848}
  {\bibfield  {journal} {\bibinfo  {journal} {Appl. Phys. Lett.}\ }\textbf
  {\bibinfo {volume} {102}},\ \bibinfo {pages} {212601} (\bibinfo {year}
  {2013})}\BibitemShut {NoStop}%
\bibitem [{\citenamefont {Rablen}\ \emph {et~al.}(2011)\citenamefont {Rablen},
  \citenamefont {Kemmler}, \citenamefont {Quaglio}, \citenamefont {Kleiner},
  \citenamefont {Koelle},\ and\ \citenamefont
  {Grigorieva}}]{PRBBitterDecoration}%
  \BibitemOpen
  \bibfield  {author} {\bibinfo {author} {\bibfnamefont {S.}~\bibnamefont
  {Rablen}}, \bibinfo {author} {\bibfnamefont {M.}~\bibnamefont {Kemmler}},
  \bibinfo {author} {\bibfnamefont {T.}~\bibnamefont {Quaglio}}, \bibinfo
  {author} {\bibfnamefont {R.}~\bibnamefont {Kleiner}}, \bibinfo {author}
  {\bibfnamefont {D.}~\bibnamefont {Koelle}}, \ and\ \bibinfo {author}
  {\bibfnamefont {I.~V.}\ \bibnamefont {Grigorieva}},\ }\href {\doibase
  10.1103/PhysRevB.84.184520} {\bibfield  {journal} {\bibinfo  {journal} {Phys.
  Rev. B}\ }\textbf {\bibinfo {volume} {84}},\ \bibinfo {pages} {184520}
  (\bibinfo {year} {2011})}\BibitemShut {NoStop}%
\bibitem [{\citenamefont {Latimer}\ \emph {et~al.}(2013)\citenamefont
  {Latimer}, \citenamefont {Berdiyorov}, \citenamefont {Xiao}, \citenamefont
  {Peeters},\ and\ \citenamefont {Kwok}}]{VortexIcePRLXiao}%
  \BibitemOpen
  \bibfield  {author} {\bibinfo {author} {\bibfnamefont {M.~L.}\ \bibnamefont
  {Latimer}}, \bibinfo {author} {\bibfnamefont {G.~R.}\ \bibnamefont
  {Berdiyorov}}, \bibinfo {author} {\bibfnamefont {Z.~L.}\ \bibnamefont
  {Xiao}}, \bibinfo {author} {\bibfnamefont {F.~M.}\ \bibnamefont {Peeters}}, \
  and\ \bibinfo {author} {\bibfnamefont {W.~K.}\ \bibnamefont {Kwok}},\ }\href
  {\doibase 10.1103/PhysRevLett.111.067001} {\bibfield  {journal} {\bibinfo
  {journal} {Phys. Rev. Lett.}\ }\textbf {\bibinfo {volume} {111}},\ \bibinfo
  {pages} {067001} (\bibinfo {year} {2013})}\BibitemShut {NoStop}%
\bibitem [{\citenamefont {Wang}\ \emph {et~al.}(2013)\citenamefont {Wang},
  \citenamefont {Latimer}, \citenamefont {Xiao}, \citenamefont {Divan},
  \citenamefont {Ocola}, \citenamefont {Crabtree},\ and\ \citenamefont
  {Kwok}}]{ConformalWang}%
  \BibitemOpen
  \bibfield  {author} {\bibinfo {author} {\bibfnamefont {Y.~L.}\ \bibnamefont
  {Wang}}, \bibinfo {author} {\bibfnamefont {M.~L.}\ \bibnamefont {Latimer}},
  \bibinfo {author} {\bibfnamefont {Z.~L.}\ \bibnamefont {Xiao}}, \bibinfo
  {author} {\bibfnamefont {R.}~\bibnamefont {Divan}}, \bibinfo {author}
  {\bibfnamefont {L.~E.}\ \bibnamefont {Ocola}}, \bibinfo {author}
  {\bibfnamefont {G.~W.}\ \bibnamefont {Crabtree}}, \ and\ \bibinfo {author}
  {\bibfnamefont {W.~K.}\ \bibnamefont {Kwok}},\ }\href {\doibase
  10.1103/PhysRevB.87.220501} {\bibfield  {journal} {\bibinfo  {journal} {Phys.
  Rev. B}\ }\textbf {\bibinfo {volume} {87}},\ \bibinfo {pages} {220501}
  (\bibinfo {year} {2013})}\BibitemShut {NoStop}%
\bibitem [{\citenamefont {Guénon}\ \emph {et~al.}(2013)\citenamefont
  {Guénon}, \citenamefont {Rosen}, \citenamefont {Basaran},\ and\
  \citenamefont {Schuller}}]{ConformalAPL}%
  \BibitemOpen
  \bibfield  {author} {\bibinfo {author} {\bibfnamefont {S.}~\bibnamefont
  {Guénon}}, \bibinfo {author} {\bibfnamefont {Y.~J.}\ \bibnamefont {Rosen}},
  \bibinfo {author} {\bibfnamefont {A.~C.}\ \bibnamefont {Basaran}}, \ and\
  \bibinfo {author} {\bibfnamefont {I.~K.}\ \bibnamefont {Schuller}},\ }\href
  {\doibase http://dx.doi.org/10.1063/1.4811413} {\bibfield  {journal}
  {\bibinfo  {journal} {Appl. Phys. Lett.}\ }\textbf {\bibinfo {volume}
  {102}},\ \bibinfo {eid} {252602} (\bibinfo {year} {2013})}\BibitemShut
  {NoStop}%
\bibitem [{\citenamefont {Trastoy}\ \emph {et~al.}(2014)\citenamefont
  {Trastoy}, \citenamefont {Malnou}, \citenamefont {Ulysse}, \citenamefont
  {Bernard}, \citenamefont {Bergeal}, \citenamefont {Faini}, \citenamefont
  {Lesueur}, \citenamefont {Briatico},\ and\ \citenamefont
  {Villegas}}]{VortexIceNatureNanoTech}%
  \BibitemOpen
  \bibfield  {author} {\bibinfo {author} {\bibfnamefont {J.}~\bibnamefont
  {Trastoy}}, \bibinfo {author} {\bibfnamefont {M.}~\bibnamefont {Malnou}},
  \bibinfo {author} {\bibfnamefont {C.}~\bibnamefont {Ulysse}}, \bibinfo
  {author} {\bibfnamefont {R.}~\bibnamefont {Bernard}}, \bibinfo {author}
  {\bibfnamefont {N.}~\bibnamefont {Bergeal}}, \bibinfo {author} {\bibfnamefont
  {G.}~\bibnamefont {Faini}}, \bibinfo {author} {\bibfnamefont
  {J.}~\bibnamefont {Lesueur}}, \bibinfo {author} {\bibfnamefont
  {J.}~\bibnamefont {Briatico}}, \ and\ \bibinfo {author} {\bibfnamefont
  {J.~E.}\ \bibnamefont {Villegas}},\ }\href@noop {} {\bibfield  {journal}
  {\bibinfo  {journal} {Nat Nano}\ }\textbf {\bibinfo {volume} {9}},\ \bibinfo
  {pages} {710} (\bibinfo {year} {2014})}\BibitemShut {NoStop}%
\bibitem [{\citenamefont {Swiecicki}\ \emph {et~al.}(2012)\citenamefont
  {Swiecicki}, \citenamefont {Ulysse}, \citenamefont {Wolf}, \citenamefont
  {Bernard}, \citenamefont {Bergeal}, \citenamefont {Briatico}, \citenamefont
  {Faini}, \citenamefont {Lesueur},\ and\ \citenamefont
  {Villegas}}]{PRB2012MaskIrradiationPeroidic}%
  \BibitemOpen
  \bibfield  {author} {\bibinfo {author} {\bibfnamefont {I.}~\bibnamefont
  {Swiecicki}}, \bibinfo {author} {\bibfnamefont {C.}~\bibnamefont {Ulysse}},
  \bibinfo {author} {\bibfnamefont {T.}~\bibnamefont {Wolf}}, \bibinfo {author}
  {\bibfnamefont {R.}~\bibnamefont {Bernard}}, \bibinfo {author} {\bibfnamefont
  {N.}~\bibnamefont {Bergeal}}, \bibinfo {author} {\bibfnamefont
  {J.}~\bibnamefont {Briatico}}, \bibinfo {author} {\bibfnamefont
  {G.}~\bibnamefont {Faini}}, \bibinfo {author} {\bibfnamefont
  {J.}~\bibnamefont {Lesueur}}, \ and\ \bibinfo {author} {\bibfnamefont
  {J.~E.}\ \bibnamefont {Villegas}},\ }\href {\doibase
  10.1103/PhysRevB.85.224502} {\bibfield  {journal} {\bibinfo  {journal} {Phys.
  Rev. B}\ }\textbf {\bibinfo {volume} {85}},\ \bibinfo {pages} {224502}
  (\bibinfo {year} {2012})}\BibitemShut {NoStop}%
\bibitem [{\citenamefont {Fang}\ \emph {et~al.}(2013)\citenamefont {Fang},
  \citenamefont {Jia}, \citenamefont {Mishra}, \citenamefont {Chaparro},
  \citenamefont {Vlasko-Vlasov}, \citenamefont {Koshelev}, \citenamefont
  {Welp}, \citenamefont {Crabtree}, \citenamefont {Zhu}, \citenamefont
  {Zhigadlo}, \citenamefont {Katrych}, \citenamefont {Karpinski},\ and\
  \citenamefont {Kwok}}]{Fang2013Fef16}%
  \BibitemOpen
  \bibfield  {author} {\bibinfo {author} {\bibfnamefont {L.}~\bibnamefont
  {Fang}}, \bibinfo {author} {\bibfnamefont {Y.}~\bibnamefont {Jia}}, \bibinfo
  {author} {\bibfnamefont {V.}~\bibnamefont {Mishra}}, \bibinfo {author}
  {\bibfnamefont {C.}~\bibnamefont {Chaparro}}, \bibinfo {author}
  {\bibfnamefont {V.~K.}\ \bibnamefont {Vlasko-Vlasov}}, \bibinfo {author}
  {\bibfnamefont {A.~E.}\ \bibnamefont {Koshelev}}, \bibinfo {author}
  {\bibfnamefont {U.}~\bibnamefont {Welp}}, \bibinfo {author} {\bibfnamefont
  {G.~W.}\ \bibnamefont {Crabtree}}, \bibinfo {author} {\bibfnamefont
  {S.}~\bibnamefont {Zhu}}, \bibinfo {author} {\bibfnamefont {N.~D.}\
  \bibnamefont {Zhigadlo}}, \bibinfo {author} {\bibfnamefont {S.}~\bibnamefont
  {Katrych}}, \bibinfo {author} {\bibfnamefont {J.}~\bibnamefont {Karpinski}},
  \ and\ \bibinfo {author} {\bibfnamefont {W.~K.}\ \bibnamefont {Kwok}},\
  }\href@noop {} {\bibfield  {journal} {\bibinfo  {journal} {Nat Commun}\
  }\textbf {\bibinfo {volume} {4}},\ \bibinfo {pages} {2655} (\bibinfo {year}
  {2013})}\BibitemShut {NoStop}%
\bibitem [{\citenamefont {Miura}\ \emph {et~al.}(2013)\citenamefont {Miura},
  \citenamefont {Maiorov}, \citenamefont {Kato}, \citenamefont {Shimode},
  \citenamefont {Wada}, \citenamefont {Adachi},\ and\ \citenamefont
  {Tanabe}}]{Miura2013Ref17}%
  \BibitemOpen
  \bibfield  {author} {\bibinfo {author} {\bibfnamefont {M.}~\bibnamefont
  {Miura}}, \bibinfo {author} {\bibfnamefont {B.}~\bibnamefont {Maiorov}},
  \bibinfo {author} {\bibfnamefont {T.}~\bibnamefont {Kato}}, \bibinfo {author}
  {\bibfnamefont {T.}~\bibnamefont {Shimode}}, \bibinfo {author} {\bibfnamefont
  {K.}~\bibnamefont {Wada}}, \bibinfo {author} {\bibfnamefont {S.}~\bibnamefont
  {Adachi}}, \ and\ \bibinfo {author} {\bibfnamefont {K.}~\bibnamefont
  {Tanabe}},\ }\href@noop {} {\bibfield  {journal} {\bibinfo  {journal} {Nat
  Commun}\ }\textbf {\bibinfo {volume} {4}},\ \bibinfo {pages} {2499} (\bibinfo
  {year} {2013})}\BibitemShut {NoStop}%
\bibitem [{\citenamefont {Baert}\ \emph {et~al.}(1995)\citenamefont {Baert},
  \citenamefont {Metlushko}, \citenamefont {Jonckheere}, \citenamefont
  {Moshchalkov},\ and\ \citenamefont
  {Bruynseraede}}]{PRL1995SquareMagnetization}%
  \BibitemOpen
  \bibfield  {author} {\bibinfo {author} {\bibfnamefont {M.}~\bibnamefont
  {Baert}}, \bibinfo {author} {\bibfnamefont {V.~V.}\ \bibnamefont
  {Metlushko}}, \bibinfo {author} {\bibfnamefont {R.}~\bibnamefont
  {Jonckheere}}, \bibinfo {author} {\bibfnamefont {V.~V.}\ \bibnamefont
  {Moshchalkov}}, \ and\ \bibinfo {author} {\bibfnamefont {Y.}~\bibnamefont
  {Bruynseraede}},\ }\href {\doibase 10.1103/PhysRevLett.74.3269} {\bibfield
  {journal} {\bibinfo  {journal} {Phys. Rev. Lett.}\ }\textbf {\bibinfo
  {volume} {74}},\ \bibinfo {pages} {3269} (\bibinfo {year}
  {1995})}\BibitemShut {NoStop}%
\bibitem [{\citenamefont {Berdiyorov}\ \emph
  {et~al.}(2006{\natexlab{a}})\citenamefont {Berdiyorov}, \citenamefont
  {Milo\ifmmode \check{s}\else \v{s}\fi{}evi\ifmmode~\acute{c}\else
  \'{c}\fi{}},\ and\ \citenamefont {Peeters}}]{New21}%
  \BibitemOpen
  \bibfield  {author} {\bibinfo {author} {\bibfnamefont {G.~R.}\ \bibnamefont
  {Berdiyorov}}, \bibinfo {author} {\bibfnamefont {M.~V.}\ \bibnamefont
  {Milo\ifmmode \check{s}\else \v{s}\fi{}evi\ifmmode~\acute{c}\else
  \'{c}\fi{}}}, \ and\ \bibinfo {author} {\bibfnamefont {F.~M.}\ \bibnamefont
  {Peeters}},\ }\href {\doibase 10.1103/PhysRevLett.96.207001} {\bibfield
  {journal} {\bibinfo  {journal} {Phys. Rev. Lett.}\ }\textbf {\bibinfo
  {volume} {96}},\ \bibinfo {pages} {207001} (\bibinfo {year}
  {2006}{\natexlab{a}})}\BibitemShut {NoStop}%
\bibitem [{\citenamefont {Reichhardt}\ \emph {et~al.}(1998)\citenamefont
  {Reichhardt}, \citenamefont {Olson},\ and\ \citenamefont
  {Nori}}]{PRB1998PeroidReichhardt}%
  \BibitemOpen
  \bibfield  {author} {\bibinfo {author} {\bibfnamefont {C.}~\bibnamefont
  {Reichhardt}}, \bibinfo {author} {\bibfnamefont {C.~J.}\ \bibnamefont
  {Olson}}, \ and\ \bibinfo {author} {\bibfnamefont {F.}~\bibnamefont {Nori}},\
  }\href {\doibase 10.1103/PhysRevB.57.7937} {\bibfield  {journal} {\bibinfo
  {journal} {Phys. Rev. B}\ }\textbf {\bibinfo {volume} {57}},\ \bibinfo
  {pages} {7937} (\bibinfo {year} {1998})}\BibitemShut {NoStop}%
\bibitem [{\citenamefont {Reichhardt}\ \emph {et~al.}(2001)\citenamefont
  {Reichhardt}, \citenamefont {Zim\'anyi}, \citenamefont {Scalettar},
  \citenamefont {Hoffmann},\ and\ \citenamefont
  {Schuller}}]{PRB2001PeroidReichhardt}%
  \BibitemOpen
  \bibfield  {author} {\bibinfo {author} {\bibfnamefont {C.}~\bibnamefont
  {Reichhardt}}, \bibinfo {author} {\bibfnamefont {G.~T.}\ \bibnamefont
  {Zim\'anyi}}, \bibinfo {author} {\bibfnamefont {R.~T.}\ \bibnamefont
  {Scalettar}}, \bibinfo {author} {\bibfnamefont {A.}~\bibnamefont {Hoffmann}},
  \ and\ \bibinfo {author} {\bibfnamefont {I.~K.}\ \bibnamefont {Schuller}},\
  }\href {\doibase 10.1103/PhysRevB.64.052503} {\bibfield  {journal} {\bibinfo
  {journal} {Phys. Rev. B}\ }\textbf {\bibinfo {volume} {64}},\ \bibinfo
  {pages} {052503} (\bibinfo {year} {2001})}\BibitemShut {NoStop}%
\bibitem [{\citenamefont {Reichhardt}\ and\ \citenamefont
  {Gr\o{}nbech-Jensen}(2001)}]{PRB2001PeroidicReichhardt2}%
  \BibitemOpen
  \bibfield  {author} {\bibinfo {author} {\bibfnamefont {C.}~\bibnamefont
  {Reichhardt}}\ and\ \bibinfo {author} {\bibfnamefont {N.}~\bibnamefont
  {Gr\o{}nbech-Jensen}},\ }\href {\doibase 10.1103/PhysRevB.63.054510}
  {\bibfield  {journal} {\bibinfo  {journal} {Phys. Rev. B}\ }\textbf {\bibinfo
  {volume} {63}},\ \bibinfo {pages} {054510} (\bibinfo {year}
  {2001})}\BibitemShut {NoStop}%
\bibitem [{\citenamefont {Misko}\ \emph {et~al.}(2005)\citenamefont {Misko},
  \citenamefont {Savel'ev},\ and\ \citenamefont {Nori}}]{Ref20}%
  \BibitemOpen
  \bibfield  {author} {\bibinfo {author} {\bibfnamefont {V.}~\bibnamefont
  {Misko}}, \bibinfo {author} {\bibfnamefont {S.}~\bibnamefont {Savel'ev}}, \
  and\ \bibinfo {author} {\bibfnamefont {F.}~\bibnamefont {Nori}},\ }\href
  {\doibase 10.1103/PhysRevLett.95.177007} {\bibfield  {journal} {\bibinfo
  {journal} {Phys. Rev. Lett.}\ }\textbf {\bibinfo {volume} {95}},\ \bibinfo
  {pages} {177007} (\bibinfo {year} {2005})}\BibitemShut {NoStop}%
\bibitem [{\citenamefont {Misko}\ \emph {et~al.}(2006)\citenamefont {Misko},
  \citenamefont {Savel'ev},\ and\ \citenamefont {Nori}}]{Ref21}%
  \BibitemOpen
  \bibfield  {author} {\bibinfo {author} {\bibfnamefont {V.~R.}\ \bibnamefont
  {Misko}}, \bibinfo {author} {\bibfnamefont {S.}~\bibnamefont {Savel'ev}}, \
  and\ \bibinfo {author} {\bibfnamefont {F.}~\bibnamefont {Nori}},\ }\href
  {\doibase 10.1103/PhysRevB.74.024522} {\bibfield  {journal} {\bibinfo
  {journal} {Phys. Rev. B}\ }\textbf {\bibinfo {volume} {74}},\ \bibinfo
  {pages} {024522} (\bibinfo {year} {2006})}\BibitemShut {NoStop}%
\bibitem [{\citenamefont {Misko}\ \emph {et~al.}(2010)\citenamefont {Misko},
  \citenamefont {Bothner}, \citenamefont {Kemmler}, \citenamefont {Kleiner},
  \citenamefont {Koelle}, \citenamefont {Peeters},\ and\ \citenamefont
  {Nori}}]{Ref22}%
  \BibitemOpen
  \bibfield  {author} {\bibinfo {author} {\bibfnamefont {V.~R.}\ \bibnamefont
  {Misko}}, \bibinfo {author} {\bibfnamefont {D.}~\bibnamefont {Bothner}},
  \bibinfo {author} {\bibfnamefont {M.}~\bibnamefont {Kemmler}}, \bibinfo
  {author} {\bibfnamefont {R.}~\bibnamefont {Kleiner}}, \bibinfo {author}
  {\bibfnamefont {D.}~\bibnamefont {Koelle}}, \bibinfo {author} {\bibfnamefont
  {F.~M.}\ \bibnamefont {Peeters}}, \ and\ \bibinfo {author} {\bibfnamefont
  {F.}~\bibnamefont {Nori}},\ }\href {\doibase 10.1103/PhysRevB.82.184512}
  {\bibfield  {journal} {\bibinfo  {journal} {Phys. Rev. B}\ }\textbf {\bibinfo
  {volume} {82}},\ \bibinfo {pages} {184512} (\bibinfo {year}
  {2010})}\BibitemShut {NoStop}%
\bibitem [{\citenamefont {Kramer}\ \emph {et~al.}(2009)\citenamefont {Kramer},
  \citenamefont {Silhanek}, \citenamefont {Van~de Vondel}, \citenamefont
  {Raes},\ and\ \citenamefont {Moshchalkov}}]{Ref23}%
  \BibitemOpen
  \bibfield  {author} {\bibinfo {author} {\bibfnamefont {R.~B.~G.}\
  \bibnamefont {Kramer}}, \bibinfo {author} {\bibfnamefont {A.~V.}\
  \bibnamefont {Silhanek}}, \bibinfo {author} {\bibfnamefont {J.}~\bibnamefont
  {Van~de Vondel}}, \bibinfo {author} {\bibfnamefont {B.}~\bibnamefont {Raes}},
  \ and\ \bibinfo {author} {\bibfnamefont {V.~V.}\ \bibnamefont
  {Moshchalkov}},\ }\href {\doibase 10.1103/PhysRevLett.103.067007} {\bibfield
  {journal} {\bibinfo  {journal} {Phys. Rev. Lett.}\ }\textbf {\bibinfo
  {volume} {103}},\ \bibinfo {pages} {067007} (\bibinfo {year}
  {2009})}\BibitemShut {NoStop}%
\bibitem [{\citenamefont {Silhanek}\ \emph {et~al.}(2006)\citenamefont
  {Silhanek}, \citenamefont {Gillijns}, \citenamefont {Moshchalkov},
  \citenamefont {Zhu}, \citenamefont {Moonens},\ and\ \citenamefont
  {Leunissen}}]{Ref24}%
  \BibitemOpen
  \bibfield  {author} {\bibinfo {author} {\bibfnamefont {A.~V.}\ \bibnamefont
  {Silhanek}}, \bibinfo {author} {\bibfnamefont {W.}~\bibnamefont {Gillijns}},
  \bibinfo {author} {\bibfnamefont {V.~V.}\ \bibnamefont {Moshchalkov}},
  \bibinfo {author} {\bibfnamefont {B.~Y.}\ \bibnamefont {Zhu}}, \bibinfo
  {author} {\bibfnamefont {J.}~\bibnamefont {Moonens}}, \ and\ \bibinfo
  {author} {\bibfnamefont {L.~H.~A.}\ \bibnamefont {Leunissen}},\ }\href
  {\doibase 10.1063/1.2361172} {\bibfield  {journal} {\bibinfo  {journal}
  {Appl. Phys. Lett.}\ }\textbf {\bibinfo {volume} {89}},\ \bibinfo {eid}
  {152507} (\bibinfo {year} {2006})}\BibitemShut {NoStop}%
\bibitem [{\citenamefont {Misko}\ and\ \citenamefont {Nori}(2012)}]{Ref27}%
  \BibitemOpen
  \bibfield  {author} {\bibinfo {author} {\bibfnamefont {V.~R.}\ \bibnamefont
  {Misko}}\ and\ \bibinfo {author} {\bibfnamefont {F.}~\bibnamefont {Nori}},\
  }\href {\doibase 10.1103/PhysRevB.85.184506} {\bibfield  {journal} {\bibinfo
  {journal} {Phys. Rev. B}\ }\textbf {\bibinfo {volume} {85}},\ \bibinfo
  {pages} {184506} (\bibinfo {year} {2012})}\BibitemShut {NoStop}%
\bibitem [{\citenamefont {Wu}\ \emph {et~al.}(2005)\citenamefont {Wu},
  \citenamefont {Wang}, \citenamefont {Horng}, \citenamefont {Wu},\ and\
  \citenamefont {Yang}}]{HoneycombJAP2005}%
  \BibitemOpen
  \bibfield  {author} {\bibinfo {author} {\bibfnamefont {T.~C.}\ \bibnamefont
  {Wu}}, \bibinfo {author} {\bibfnamefont {J.~C.}\ \bibnamefont {Wang}},
  \bibinfo {author} {\bibfnamefont {L.}~\bibnamefont {Horng}}, \bibinfo
  {author} {\bibfnamefont {J.~C.}\ \bibnamefont {Wu}}, \ and\ \bibinfo {author}
  {\bibfnamefont {T.~J.}\ \bibnamefont {Yang}},\ }\href {\doibase
  http://dx.doi.org/10.1063/1.1849511} {\bibfield  {journal} {\bibinfo
  {journal} {J. Appl. Phys.}\ }\textbf {\bibinfo {volume} {97}},\ \bibinfo
  {eid} {10B102} (\bibinfo {year} {2005})}\BibitemShut {NoStop}%
\bibitem [{\citenamefont {Reichhardt}\ and\ \citenamefont
  {Reichhardt}(2007{\natexlab{a}})}]{Honeycomb2007Reichhardt}%
  \BibitemOpen
  \bibfield  {author} {\bibinfo {author} {\bibfnamefont {C.}~\bibnamefont
  {Reichhardt}}\ and\ \bibinfo {author} {\bibfnamefont {C.~J.~O.}\ \bibnamefont
  {Reichhardt}},\ }\href {\doibase 10.1103/PhysRevB.76.064523} {\bibfield
  {journal} {\bibinfo  {journal} {Phys. Rev. B}\ }\textbf {\bibinfo {volume}
  {76}},\ \bibinfo {pages} {064523} (\bibinfo {year}
  {2007}{\natexlab{a}})}\BibitemShut {NoStop}%
\bibitem [{\citenamefont {Reichhardt}\ and\ \citenamefont
  {Reichhardt}(2008)}]{Honeycomb2008Reichhardt}%
  \BibitemOpen
  \bibfield  {author} {\bibinfo {author} {\bibfnamefont {C.}~\bibnamefont
  {Reichhardt}}\ and\ \bibinfo {author} {\bibfnamefont {C.~J.~O.}\ \bibnamefont
  {Reichhardt}},\ }\href {\doibase 10.1103/PhysRevLett.100.167002} {\bibfield
  {journal} {\bibinfo  {journal} {Phys. Rev. Lett.}\ }\textbf {\bibinfo
  {volume} {100}},\ \bibinfo {pages} {167002} (\bibinfo {year}
  {2008})}\BibitemShut {NoStop}%
\bibitem [{\citenamefont {Latimer}\ \emph {et~al.}(2012)\citenamefont
  {Latimer}, \citenamefont {Berdiyorov}, \citenamefont {Xiao}, \citenamefont
  {Kwok},\ and\ \citenamefont {Peeters}}]{Honeycomb2012Xiao}%
  \BibitemOpen
  \bibfield  {author} {\bibinfo {author} {\bibfnamefont {M.~L.}\ \bibnamefont
  {Latimer}}, \bibinfo {author} {\bibfnamefont {G.~R.}\ \bibnamefont
  {Berdiyorov}}, \bibinfo {author} {\bibfnamefont {Z.~L.}\ \bibnamefont
  {Xiao}}, \bibinfo {author} {\bibfnamefont {W.~K.}\ \bibnamefont {Kwok}}, \
  and\ \bibinfo {author} {\bibfnamefont {F.~M.}\ \bibnamefont {Peeters}},\
  }\href {\doibase 10.1103/PhysRevB.85.012505} {\bibfield  {journal} {\bibinfo
  {journal} {Phys. Rev. B}\ }\textbf {\bibinfo {volume} {85}},\ \bibinfo
  {pages} {012505} (\bibinfo {year} {2012})}\BibitemShut {NoStop}%
\bibitem [{\citenamefont {Kemmler}\ \emph {et~al.}(2009)\citenamefont
  {Kemmler}, \citenamefont {Bothner}, \citenamefont {Ilin}, \citenamefont
  {Siegel}, \citenamefont {Kleiner},\ and\ \citenamefont
  {Koelle}}]{DisorderInOrder2009Kemmler}%
  \BibitemOpen
  \bibfield  {author} {\bibinfo {author} {\bibfnamefont {M.}~\bibnamefont
  {Kemmler}}, \bibinfo {author} {\bibfnamefont {D.}~\bibnamefont {Bothner}},
  \bibinfo {author} {\bibfnamefont {K.}~\bibnamefont {Ilin}}, \bibinfo {author}
  {\bibfnamefont {M.}~\bibnamefont {Siegel}}, \bibinfo {author} {\bibfnamefont
  {R.}~\bibnamefont {Kleiner}}, \ and\ \bibinfo {author} {\bibfnamefont
  {D.}~\bibnamefont {Koelle}},\ }\href {\doibase 10.1103/PhysRevB.79.184509}
  {\bibfield  {journal} {\bibinfo  {journal} {Phys. Rev. B}\ }\textbf {\bibinfo
  {volume} {79}},\ \bibinfo {pages} {184509} (\bibinfo {year}
  {2009})}\BibitemShut {NoStop}%
\bibitem [{\citenamefont {Ray}\ \emph {et~al.}(2013)\citenamefont {Ray},
  \citenamefont {Olson~Reichhardt}, \citenamefont {Jank\'o},\ and\
  \citenamefont {Reichhardt}}]{ConformalRayPRL}%
  \BibitemOpen
  \bibfield  {author} {\bibinfo {author} {\bibfnamefont {D.}~\bibnamefont
  {Ray}}, \bibinfo {author} {\bibfnamefont {C.~J.}\ \bibnamefont
  {Olson~Reichhardt}}, \bibinfo {author} {\bibfnamefont {B.}~\bibnamefont
  {Jank\'o}}, \ and\ \bibinfo {author} {\bibfnamefont {C.}~\bibnamefont
  {Reichhardt}},\ }\href {\doibase 10.1103/PhysRevLett.110.267001} {\bibfield
  {journal} {\bibinfo  {journal} {Phys. Rev. Lett.}\ }\textbf {\bibinfo
  {volume} {110}},\ \bibinfo {pages} {267001} (\bibinfo {year}
  {2013})}\BibitemShut {NoStop}%
\bibitem [{\citenamefont {Ray}\ \emph {et~al.}(2014)\citenamefont {Ray},
  \citenamefont {Reichhardt},\ and\ \citenamefont
  {Reichhardt}}]{ConformalRayPRB}%
  \BibitemOpen
  \bibfield  {author} {\bibinfo {author} {\bibfnamefont {D.}~\bibnamefont
  {Ray}}, \bibinfo {author} {\bibfnamefont {C.}~\bibnamefont {Reichhardt}}, \
  and\ \bibinfo {author} {\bibfnamefont {C.~J.~O.}\ \bibnamefont
  {Reichhardt}},\ }\href {\doibase 10.1103/PhysRevB.90.094502} {\bibfield
  {journal} {\bibinfo  {journal} {Phys. Rev. B}\ }\textbf {\bibinfo {volume}
  {90}},\ \bibinfo {pages} {094502} (\bibinfo {year} {2014})}\BibitemShut
  {NoStop}%
\bibitem [{\citenamefont {MacManus-Driscoll}\ \emph {et~al.}(2004)\citenamefont
  {MacManus-Driscoll}, \citenamefont {Foltyn}, \citenamefont {Jia},
  \citenamefont {Wang}, \citenamefont {Serquis}, \citenamefont {Civale},
  \citenamefont {Maiorov}, \citenamefont {Hawley}, \citenamefont {Maley},\ and\
  \citenamefont {Peterson}}]{MacManus-Driscoll2004}%
  \BibitemOpen
  \bibfield  {author} {\bibinfo {author} {\bibfnamefont {J.~L.}\ \bibnamefont
  {MacManus-Driscoll}}, \bibinfo {author} {\bibfnamefont {S.~R.}\ \bibnamefont
  {Foltyn}}, \bibinfo {author} {\bibfnamefont {Q.~X.}\ \bibnamefont {Jia}},
  \bibinfo {author} {\bibfnamefont {H.}~\bibnamefont {Wang}}, \bibinfo {author}
  {\bibfnamefont {A.}~\bibnamefont {Serquis}}, \bibinfo {author} {\bibfnamefont
  {L.}~\bibnamefont {Civale}}, \bibinfo {author} {\bibfnamefont
  {B.}~\bibnamefont {Maiorov}}, \bibinfo {author} {\bibfnamefont {M.~E.}\
  \bibnamefont {Hawley}}, \bibinfo {author} {\bibfnamefont {M.~P.}\
  \bibnamefont {Maley}}, \ and\ \bibinfo {author} {\bibfnamefont {D.~E.}\
  \bibnamefont {Peterson}},\ }\href {http://dx.doi.org/10.1038/nmat1156}
  {\bibfield  {journal} {\bibinfo  {journal} {Nat Mater}\ }\textbf {\bibinfo
  {volume} {3}},\ \bibinfo {pages} {439} (\bibinfo {year} {2004})}\BibitemShut
  {NoStop}%
\bibitem [{\citenamefont {Krusin-Elbaum}\ \emph {et~al.}(2000)\citenamefont
  {Krusin-Elbaum}, \citenamefont {Blatter}, \citenamefont {Thompson},
  \citenamefont {Ullmann},\ and\ \citenamefont {Chu}}]{Krusin-Elbaum2000}%
  \BibitemOpen
  \bibfield  {author} {\bibinfo {author} {\bibfnamefont {L.}~\bibnamefont
  {Krusin-Elbaum}}, \bibinfo {author} {\bibfnamefont {G.}~\bibnamefont
  {Blatter}}, \bibinfo {author} {\bibfnamefont {J.}~\bibnamefont {Thompson}},
  \bibinfo {author} {\bibfnamefont {J.}~\bibnamefont {Ullmann}}, \ and\
  \bibinfo {author} {\bibfnamefont {C.}~\bibnamefont {Chu}},\ }\href
  {http://www.sciencedirect.com/science/article/pii/S092145340000160X}
  {\bibfield  {journal} {\bibinfo  {journal} {Physica C: Superconductivity}\
  }\textbf {\bibinfo {volume} {335}},\ \bibinfo {pages} {144} (\bibinfo {year}
  {2000})}\BibitemShut {NoStop}%
\bibitem [{\citenamefont {Rosen}\ \emph {et~al.}(2010)\citenamefont {Rosen},
  \citenamefont {Sharoni},\ and\ \citenamefont
  {Schuller}}]{PhysRevB.82.014509}%
  \BibitemOpen
  \bibfield  {author} {\bibinfo {author} {\bibfnamefont {Y.~J.}\ \bibnamefont
  {Rosen}}, \bibinfo {author} {\bibfnamefont {A.}~\bibnamefont {Sharoni}}, \
  and\ \bibinfo {author} {\bibfnamefont {I.~K.}\ \bibnamefont {Schuller}},\
  }\href {\doibase 10.1103/PhysRevB.82.014509} {\bibfield  {journal} {\bibinfo
  {journal} {Phys. Rev. B}\ }\textbf {\bibinfo {volume} {82}},\ \bibinfo
  {pages} {014509} (\bibinfo {year} {2010})}\BibitemShut {NoStop}%
\bibitem [{\citenamefont {Reichhardt}\ \emph {et~al.}(1996)\citenamefont
  {Reichhardt}, \citenamefont {Olson}, \citenamefont {Groth}, \citenamefont
  {Field},\ and\ \citenamefont {Nori}}]{Random1996Reichhardt}%
  \BibitemOpen
  \bibfield  {author} {\bibinfo {author} {\bibfnamefont {C.}~\bibnamefont
  {Reichhardt}}, \bibinfo {author} {\bibfnamefont {C.~J.}\ \bibnamefont
  {Olson}}, \bibinfo {author} {\bibfnamefont {J.}~\bibnamefont {Groth}},
  \bibinfo {author} {\bibfnamefont {S.}~\bibnamefont {Field}}, \ and\ \bibinfo
  {author} {\bibfnamefont {F.}~\bibnamefont {Nori}},\ }\href {\doibase
  10.1103/PhysRevB.53.R8898} {\bibfield  {journal} {\bibinfo  {journal} {Phys.
  Rev. B}\ }\textbf {\bibinfo {volume} {53}},\ \bibinfo {pages} {R8898}
  (\bibinfo {year} {1996})}\BibitemShut {NoStop}%
\bibitem [{\citenamefont {Liang}\ \emph {et~al.}(2010)\citenamefont {Liang},
  \citenamefont {Kunchur}, \citenamefont {Hua},\ and\ \citenamefont
  {Xiao}}]{Ref35}%
  \BibitemOpen
  \bibfield  {author} {\bibinfo {author} {\bibfnamefont {M.}~\bibnamefont
  {Liang}}, \bibinfo {author} {\bibfnamefont {M.~N.}\ \bibnamefont {Kunchur}},
  \bibinfo {author} {\bibfnamefont {J.}~\bibnamefont {Hua}}, \ and\ \bibinfo
  {author} {\bibfnamefont {Z.}~\bibnamefont {Xiao}},\ }\href {\doibase
  10.1103/PhysRevB.82.064502} {\bibfield  {journal} {\bibinfo  {journal} {Phys.
  Rev. B}\ }\textbf {\bibinfo {volume} {82}},\ \bibinfo {pages} {064502}
  (\bibinfo {year} {2010})}\BibitemShut {NoStop}%
\bibitem [{\citenamefont {Patel}\ \emph {et~al.}(2007)\citenamefont {Patel},
  \citenamefont {Xiao}, \citenamefont {Hua}, \citenamefont {Xu}, \citenamefont
  {Rosenmann}, \citenamefont {Novosad}, \citenamefont {Pearson}, \citenamefont
  {Welp}, \citenamefont {Kwok},\ and\ \citenamefont {Crabtree}}]{Ref36}%
  \BibitemOpen
  \bibfield  {author} {\bibinfo {author} {\bibfnamefont {U.}~\bibnamefont
  {Patel}}, \bibinfo {author} {\bibfnamefont {Z.~L.}\ \bibnamefont {Xiao}},
  \bibinfo {author} {\bibfnamefont {J.}~\bibnamefont {Hua}}, \bibinfo {author}
  {\bibfnamefont {T.}~\bibnamefont {Xu}}, \bibinfo {author} {\bibfnamefont
  {D.}~\bibnamefont {Rosenmann}}, \bibinfo {author} {\bibfnamefont
  {V.}~\bibnamefont {Novosad}}, \bibinfo {author} {\bibfnamefont
  {J.}~\bibnamefont {Pearson}}, \bibinfo {author} {\bibfnamefont
  {U.}~\bibnamefont {Welp}}, \bibinfo {author} {\bibfnamefont {W.~K.}\
  \bibnamefont {Kwok}}, \ and\ \bibinfo {author} {\bibfnamefont {G.~W.}\
  \bibnamefont {Crabtree}},\ }\href {\doibase 10.1103/PhysRevB.76.020508}
  {\bibfield  {journal} {\bibinfo  {journal} {Phys. Rev. B}\ }\textbf {\bibinfo
  {volume} {76}},\ \bibinfo {pages} {020508} (\bibinfo {year}
  {2007})}\BibitemShut {NoStop}%
\bibitem [{\citenamefont {Sochnikov}\ \emph {et~al.}(2010)\citenamefont
  {Sochnikov}, \citenamefont {Shaulov}, \citenamefont {Yeshurun}, \citenamefont
  {Logvenov},\ and\ \citenamefont {Bozovic}}]{Sochnikov2010}%
  \BibitemOpen
  \bibfield  {author} {\bibinfo {author} {\bibfnamefont {I.}~\bibnamefont
  {Sochnikov}}, \bibinfo {author} {\bibfnamefont {A.}~\bibnamefont {Shaulov}},
  \bibinfo {author} {\bibfnamefont {Y.}~\bibnamefont {Yeshurun}}, \bibinfo
  {author} {\bibfnamefont {G.}~\bibnamefont {Logvenov}}, \ and\ \bibinfo
  {author} {\bibfnamefont {I.}~\bibnamefont {Bozovic}},\ }\href@noop {}
  {\bibfield  {journal} {\bibinfo  {journal} {Nat Nano}\ }\textbf {\bibinfo
  {volume} {5}},\ \bibinfo {pages} {516} (\bibinfo {year} {2010})}\BibitemShut
  {NoStop}%
\bibitem [{\citenamefont {Berdiyorov}\ \emph {et~al.}(2012)\citenamefont
  {Berdiyorov}, \citenamefont {Milo\ifmmode \check{s}\else
  \v{s}\fi{}evi\ifmmode~\acute{c}\else \'{c}\fi{}}, \citenamefont {Latimer},
  \citenamefont {Xiao}, \citenamefont {Kwok},\ and\ \citenamefont
  {Peeters}}]{New40}%
  \BibitemOpen
  \bibfield  {author} {\bibinfo {author} {\bibfnamefont {G.~R.}\ \bibnamefont
  {Berdiyorov}}, \bibinfo {author} {\bibfnamefont {M.~V.}\ \bibnamefont
  {Milo\ifmmode \check{s}\else \v{s}\fi{}evi\ifmmode~\acute{c}\else
  \'{c}\fi{}}}, \bibinfo {author} {\bibfnamefont {M.~L.}\ \bibnamefont
  {Latimer}}, \bibinfo {author} {\bibfnamefont {Z.~L.}\ \bibnamefont {Xiao}},
  \bibinfo {author} {\bibfnamefont {W.~K.}\ \bibnamefont {Kwok}}, \ and\
  \bibinfo {author} {\bibfnamefont {F.~M.}\ \bibnamefont {Peeters}},\ }\href
  {\doibase 10.1103/PhysRevLett.109.057004} {\bibfield  {journal} {\bibinfo
  {journal} {Phys. Rev. Lett.}\ }\textbf {\bibinfo {volume} {109}},\ \bibinfo
  {pages} {057004} (\bibinfo {year} {2012})}\BibitemShut {NoStop}%
\bibitem [{\citenamefont {Reichhardt}\ and\ \citenamefont
  {Olson~Reichhardt}(2009)}]{Ref37}%
  \BibitemOpen
  \bibfield  {author} {\bibinfo {author} {\bibfnamefont {C.}~\bibnamefont
  {Reichhardt}}\ and\ \bibinfo {author} {\bibfnamefont {C.~J.}\ \bibnamefont
  {Olson~Reichhardt}},\ }\href {\doibase 10.1103/PhysRevB.79.134501} {\bibfield
   {journal} {\bibinfo  {journal} {Phys. Rev. B}\ }\textbf {\bibinfo {volume}
  {79}},\ \bibinfo {pages} {134501} (\bibinfo {year} {2009})}\BibitemShut
  {NoStop}%
\bibitem [{\citenamefont {Berdiyorov}\ \emph
  {et~al.}(2006{\natexlab{b}})\citenamefont {Berdiyorov}, \citenamefont
  {Milosevic},\ and\ \citenamefont {Peeters}}]{New44}%
  \BibitemOpen
  \bibfield  {author} {\bibinfo {author} {\bibfnamefont {G.~R.}\ \bibnamefont
  {Berdiyorov}}, \bibinfo {author} {\bibfnamefont {M.~V.}\ \bibnamefont
  {Milosevic}}, \ and\ \bibinfo {author} {\bibfnamefont {F.~M.}\ \bibnamefont
  {Peeters}},\ }\href@noop {} {\bibfield  {journal} {\bibinfo  {journal} {EPL
  (Europhysics Letters)}\ }\textbf {\bibinfo {volume} {74}},\ \bibinfo {pages}
  {493} (\bibinfo {year} {2006}{\natexlab{b}})}\BibitemShut {NoStop}%
\bibitem [{\citenamefont {Aurenhammer}(1991)}]{VoronoiDiagram}%
  \BibitemOpen
  \bibfield  {author} {\bibinfo {author} {\bibfnamefont {F.}~\bibnamefont
  {Aurenhammer}},\ }\href {\doibase 10.1145/116873.116880} {\bibfield
  {journal} {\bibinfo  {journal} {ACM Comput. Surv.}\ }\textbf {\bibinfo
  {volume} {23}},\ \bibinfo {pages} {345} (\bibinfo {year} {1991})}\BibitemShut
  {NoStop}%
\bibitem [{\citenamefont {Reichhardt}\ and\ \citenamefont
  {Reichhardt}(2007{\natexlab{b}})}]{Disorder2007Reichhardt}%
  \BibitemOpen
  \bibfield  {author} {\bibinfo {author} {\bibfnamefont {C.}~\bibnamefont
  {Reichhardt}}\ and\ \bibinfo {author} {\bibfnamefont {C.~J.~O.}\ \bibnamefont
  {Reichhardt}},\ }\href {\doibase 10.1103/PhysRevB.76.094512} {\bibfield
  {journal} {\bibinfo  {journal} {Phys. Rev. B}\ }\textbf {\bibinfo {volume}
  {76}},\ \bibinfo {pages} {094512} (\bibinfo {year}
  {2007}{\natexlab{b}})}\BibitemShut {NoStop}%
\bibitem [{\citenamefont {Mangold}\ \emph {et~al.}(2003)\citenamefont
  {Mangold}, \citenamefont {Leiderer},\ and\ \citenamefont
  {Bechinger}}]{Colloidal}%
  \BibitemOpen
  \bibfield  {author} {\bibinfo {author} {\bibfnamefont {K.}~\bibnamefont
  {Mangold}}, \bibinfo {author} {\bibfnamefont {P.}~\bibnamefont {Leiderer}}, \
  and\ \bibinfo {author} {\bibfnamefont {C.}~\bibnamefont {Bechinger}},\ }\href
  {\doibase 10.1103/PhysRevLett.90.158302} {\bibfield  {journal} {\bibinfo
  {journal} {Phys. Rev. Lett.}\ }\textbf {\bibinfo {volume} {90}},\ \bibinfo
  {pages} {158302} (\bibinfo {year} {2003})}\BibitemShut {NoStop}%
\bibitem [{\citenamefont {Pertsinidis}\ and\ \citenamefont
  {Ling}(2001)}]{Pertsinidis2001}%
  \BibitemOpen
  \bibfield  {author} {\bibinfo {author} {\bibfnamefont {A.}~\bibnamefont
  {Pertsinidis}}\ and\ \bibinfo {author} {\bibfnamefont {X.~S.}\ \bibnamefont
  {Ling}},\ }\href@noop {} {\bibfield  {journal} {\bibinfo  {journal} {Nature}\
  }\textbf {\bibinfo {volume} {413}},\ \bibinfo {pages} {147} (\bibinfo {year}
  {2001})}\BibitemShut {NoStop}%
\bibitem [{\citenamefont {Gokhale}\ \emph {et~al.}(2014)\citenamefont
  {Gokhale}, \citenamefont {Hima~Nagamanasa}, \citenamefont {Ganapathy},\ and\
  \citenamefont {Sood}}]{Gokhale2014}%
  \BibitemOpen
  \bibfield  {author} {\bibinfo {author} {\bibfnamefont {S.}~\bibnamefont
  {Gokhale}}, \bibinfo {author} {\bibfnamefont {K.}~\bibnamefont
  {Hima~Nagamanasa}}, \bibinfo {author} {\bibfnamefont {R.}~\bibnamefont
  {Ganapathy}}, \ and\ \bibinfo {author} {\bibfnamefont {A.~K.}\ \bibnamefont
  {Sood}},\ }\href@noop {} {\bibfield  {journal} {\bibinfo  {journal} {Nat
  Commun}\ }\textbf {\bibinfo {volume} {5}},\ \bibinfo {pages} {4685} (\bibinfo
  {year} {2014})}\BibitemShut {NoStop}%
\bibitem [{\citenamefont {Tung}\ \emph {et~al.}(2006)\citenamefont {Tung},
  \citenamefont {Schweikhard},\ and\ \citenamefont {Cornell}}]{BES}%
  \BibitemOpen
  \bibfield  {author} {\bibinfo {author} {\bibfnamefont {S.}~\bibnamefont
  {Tung}}, \bibinfo {author} {\bibfnamefont {V.}~\bibnamefont {Schweikhard}}, \
  and\ \bibinfo {author} {\bibfnamefont {E.~A.}\ \bibnamefont {Cornell}},\
  }\href {\doibase 10.1103/PhysRevLett.97.240402} {\bibfield  {journal}
  {\bibinfo  {journal} {Phys. Rev. Lett.}\ }\textbf {\bibinfo {volume} {97}},\
  \bibinfo {pages} {240402} (\bibinfo {year} {2006})}\BibitemShut {NoStop}%
\bibitem [{\citenamefont {Haller}\ \emph {et~al.}(2010)\citenamefont {Haller},
  \citenamefont {Hart}, \citenamefont {Mark}, \citenamefont {Danzl},
  \citenamefont {Reichsollner}, \citenamefont {Gustavsson}, \citenamefont
  {Dalmonte}, \citenamefont {Pupillo},\ and\ \citenamefont
  {Nagerl}}]{Haller2010LuttingerLiquid}%
  \BibitemOpen
  \bibfield  {author} {\bibinfo {author} {\bibfnamefont {E.}~\bibnamefont
  {Haller}}, \bibinfo {author} {\bibfnamefont {R.}~\bibnamefont {Hart}},
  \bibinfo {author} {\bibfnamefont {M.~J.}\ \bibnamefont {Mark}}, \bibinfo
  {author} {\bibfnamefont {J.~G.}\ \bibnamefont {Danzl}}, \bibinfo {author}
  {\bibfnamefont {L.}~\bibnamefont {Reichsollner}}, \bibinfo {author}
  {\bibfnamefont {M.}~\bibnamefont {Gustavsson}}, \bibinfo {author}
  {\bibfnamefont {M.}~\bibnamefont {Dalmonte}}, \bibinfo {author}
  {\bibfnamefont {G.}~\bibnamefont {Pupillo}}, \ and\ \bibinfo {author}
  {\bibfnamefont {H.-C.}\ \bibnamefont {Nagerl}},\ }\href@noop {} {\bibfield
  {journal} {\bibinfo  {journal} {Nature}\ }\textbf {\bibinfo {volume} {466}},\
  \bibinfo {pages} {597} (\bibinfo {year} {2010})}\BibitemShut {NoStop}%
\end{thebibliography}%

\end{document}